%% file: mcintosh_v2.tex
\documentclass[usenatbib]{mn2e}
\usepackage{graphicx,times,amsmath}

\usepackage[unicode=true,bookmarks=true,bookmarksnumbered=true,bookmarksopen=true,bookmarksopenlevel=1,breaklinks=false,pdfborder={0 0 0},backref=false,colorlinks=false]{hyperref}

\input{macros}

\title[Recently quenched elliptical galaxies in the SDSS]{A new population of recently quenched elliptical galaxies in the SDSS}

\author[McIntosh et al.] 
{\parbox[t]{\textwidth}{
Daniel H.\ McIntosh$^{1,}$\thanks{E-mail: mcintoshdh@umkc.edu}, 
Cory Wagner$^{1,2}$,
Andrew Cooper$^{1,3}$, Eric F.\ Bell$^4$, Du\v{s}an Kere\v{s}$^5$, 
Frank C.\ van den Bosch$^6$, Anna Gallazzi$^{7,8}$, Tim Haines$^{1,9}$, 
Justin Mann$^{1,10}$, Anna Pasquali$^{11}$, 
Allison M.\ Christian$^{1,12}$ }
\\
\vspace*{3pt} \\
$^1$ Department of Physics \& Astronomy, University of Missouri-Kansas City, 5110 Rockhill Road, Kansas City, MO 64110, USA\\
$^2$ current address: Department of Physics, Engineering Physics \& Astronomy, Queen's University, Kingston, Ontario, K7L 3N6, Canada\\
$^3$ current address: Department of Physics \& Astronomy, University of North Carolina at Chapel Hill, Chapel Hill, NC 27599, USA\\
$^4$ Department of Astronomy, University of Michigan, Ann Arbor, MI 48109, USA\\
$^5$ Department of Physics, Center for Astrophysics and Space Science, University of California San Diego, 9500 Gilman Drive, La Jolla, CA 92093, USA\\
$^6$ Astronomy Department, Yale University, New Haven, CT 06511, USA\\
$^7$ current address: INAF-Osservatorio Astrofisico di Arcetri, Largo Enrico Fermi 5, I-50125 Firenze, Italy\\
$^8$ Dark Cosmology Centre, Niels Bohr Institute, University of Copenhagen, Denmark\\
$^9$ current address: Department of Astronomy, University of Wisconsin, Madison, WI 53706, USA\\
$^{10}$ current address: Department of Physics \& Astronomy, University of Kansas, Lawrence, KS 66045, USA\\
$^{11}$ Astronomisches Rechen Institut Zentrum f\"ur Astronomie der Universit\"at Heidelberg, M\"onchhofstr. 12 - 14, D-69120 Heidelberg, Germany\\
$^{12}$ current address: Departments of Physics and Mathematics, Massachusetts Institute of Technology, Cambridge, MA 02139, USA
}

\begin{document}

\date{{\sc Draft: } \today }
\pagerange{\pageref{firstpage}--\pageref{lastpage}} \pubyear{2013}
\maketitle
\label{firstpage}

\begin{abstract}
We use the Sloan Digital Sky Survey to investigate the properties of
massive elliptical galaxies in the local Universe ($z\leq 0.08$) that have
unusually blue optical colors.
Through careful inspection, we distinguish elliptical
from non-elliptical morphologies among a large sample of similarly blue
galaxies with high central light concentrations ($c_r\geq 2.6$).
These blue ellipticals 
comprise 3.7 per cent of all $c_r\geq 2.6$ galaxies with stellar
masses between $10^{10}$ and $10^{11}~h^{-2}~{\rm M}_{\sun}$.
Using published fiber spectra diagnostics, 
we identify a unique subset of 172 {\it non-star-forming}
ellipticals with distinctly blue $urz$ colors and young ($<3$\,Gyr)
light-weighted stellar ages.
These {\it recently quenched ellipticals} (RQEs) 
have a number density of $2.7-4.7\times 10^{-5}\,h^3\,{\rm Mpc}^{-3}$
and sufficient numbers 
above $2.5\times10^{10}~h^{-2}~{\rm M}_{\sun}$ to account for more than half
of the expected quiescent growth at late cosmic time assuming
this phase lasts 0.5\,Gyr. 
RQEs have properties that are consistent with a recent merger origin
(i.e., they are strong `first-generation' elliptical candidates), 
yet few involved
a starburst strong enough to produce an E$+$A signature. The preferred
environment of RQEs (90 per cent reside at the centers of
$<3\times 10^{12}\,h^{-1}{\rm M}_{\sun}$ groups)
agrees well with the `small group scale' predicted for
maximally efficient spiral merging on to their halo center and rules out
satellite-specific quenching processes.
The high incidence of Seyfert and LINER activity in RQEs and their plausible
descendants may heat the atmospheres of small host halos sufficiently
to maintain quenching.  
\end{abstract}


\section{Introduction}

Documenting the assembly history of massive elliptical galaxies in detail
remains an elusive problem. A central aspect of galaxy evolution over
a significant portion of cosmic time has been the build up of quiescent
(i.e., non-star-forming and red) galaxies
\citep{bell04b,brown07,faber07,brammer11}.
The growth of the red galaxy population has occurred largely above the 
characteristic mass limit of
${\rm M}_{{\rm gal,}{\star}}\geq 3\times 10^{10}~{\rm M}_{\sun}$
that broadly divides galaxies into the blue-cloud of late-type (disk-dominated)
systems and the red-sequence of early-type (elliptical, S0, and
bulge-dominated spiral) galaxies, which have been well-documented at
$z\sim0$ 
\citep{strateva01,blanton03d,kauffmann03b,baldry04}.
With the advent of better and larger surveys of distant galaxies,
the bimodality in color \citep{willmer06,brammer09,whitaker11,muzzin13b}
and structure 
\citep[i.e., the high early-type fraction on the red sequence --][]{bell04a,bell12,wuyts11b,cheung12}
is found at all redshifts out to $z\sim3$.
A general consensus has emerged in the literature to explain galaxy bimodality
and its evolution, whereby the low to moderate-mass
red sequence is fed by migrating blue-cloud galaxies that experience star 
formation (SF) quenching, and the assembly of the
most massive galaxies occurs by dissipationless (so-called `dry') merging
of pre-existing red systems
\citep[see Fig.\,10 in][for an illustrative schematic diagram of the
blue-to-red migration scenario]{faber07}. The evidence for the role of
merging in the formation of $> 10^{11}~{\rm M}_{\sun}$ galaxies is
convincing \citep{bell06a,white07,mcintosh08,skelton09,vanderwel09b,robaina10}.
What remains
difficult to constrain in this model are the variety of physical processes
at play that are needed to {\it both} govern SF {\it and} alter 
structure to maintain the high
fraction of red early-type galaxies and the bimodality in galaxy properties
at masses below $10^{11}~{\rm M}_{\sun}$.
Cosmological simulations \citep[e.g.,][]{oser10} make it clear that galaxies
experience multiple processes over their lifetimes.
To gain further insights into
galaxy evolution, the work to be done is to disentangle the complex interplay 
of processes and identify which dominate under different physical conditions,
and as a function of cosmic time. 

A host of physical processes are predicted to quench SF by either removing,
heating, or cutting off the cold gas supply necessary to fuel new star
production. Energy released from accretion on to the central massive black hole
can produce dynamic-mode AGN feedback 
in the form of powerful gas outflows 
\citep{granato04} typically associated with gas-rich
mergers \citep{dimatteo05,springel05a,hopkins06b}, or 'radio-mode'
heating of the interstellar medium in galaxies \citep{sazonov05,hopkins10a}
or of the intracluster medium (ICM)
in groups and clusters \citep{cattaneo06,croton06a,sijacki06}.
During starbursts, energetic
feedback from supernovae and stellar mass loss can also provide thermal heating,
strong winds and galactic outflows 
\citep{springel03a,martin05,cox06b,tremonti07,diamondstanic12b}. 
Stellar or supernova (SN) feedback is argued to
dominate in low-mass galaxies 
\citep[e.g., $<10^{10}~{\rm M}_{\sun}$][]{kaviraj07d}, while AGN feedback
is predicted to dominate at higher masses \citep{kaviraj07d,dimatteo08b}.
Gas exhaustion and shock heating from major
gas-rich mergers {\it without} AGN or SN feedback are predicted to at
least temporarily quench SF \citep{hopkins08b}, while
other dynamical mechanisms may reduce the efficiency of SF 
including secular bar-driven quenching
\citep[e.g.,][]{masters11a}, 
and morphological quenching in
which a large spheroidal bulge can stabilize the gas disk against fragmentation
\citep{crocker12,martig13}.
Additionally, the atmosphere of a galaxy's dark matter halo can impede SF
in a number of ways. Virial shock heating of the halo gas
creates conditions
for efficient shutdown of the hot halo gas by feedback mechanisms 
\citep[e.g.,][]{keres05,dekel06a}. Cosmological
and hydrodynamical simulations show that modest-sized halos
($\geq 10^{11}-10^{12}\,{\rm M}_{\sun}$) can become
dominated by hot gas \citep{birnboim03,keres05}. Once hot, radio-mode AGN
heating \citep[or gravitational heating for larger halos,][]{dekel08}
can maintain halo quenching of
both central and satellite galaxies \citep{gabor12}. New cold-gas accretion
on to the centers of small group or galaxy size hot halos would require
additional energetic feedback to quench subsequent central SF 
\citep{keres12}. Besides preventing gas cooling, the parent halo atmosphere
can quench orbiting satellites by either tidally stripping their hot-halo
gas resulting in a so-called `strangulation' of future SF after the existing
cold fuel is consumed
\citep{larson80,balogh00a,bekki02c,kawata08},
or rapid ($\sim 10^7$\,years) ram-pressure stripping of the cold gas reservoir 
producing a fast truncation of SF
\citep{spitzer51,gunn72,abadi99,fujita99,quilis00}.
Observational results support strangulation as the dominant quenching
mechanism for the bulk of low-redshift satellite galaxies
\citep{vandenbosch08} including those in the outskirts of galaxy clusters
\citep{lewis02}.

As with quenching, many physical processes are cited to transform late-type
disks into the variety of observed early-type galaxy (ETG)
morphologies. Foremost, the hierarchical assembly of dark matter halos
\citep{white78a}
drives galaxy merging and the formation of the spheroidal components
of galaxies \citep{kauffmann93,baugh96,cole00}. 
Numerical simulations have long shown that the violent
merging of comparable mass disk galaxies (major merging) disrupts the
rotational stellar orbits
and produces remnants with spheroidal light profiles thereby giving rise
to the ``merger hypothesis'' for the formation of elliptical galaxies
\citep{toomre72,toomre77}. Soon thereafter, modellers realized the need for 
progenitor bulge components
\citep{hernquist93d} and the dissipative effects of gas
\citep{barnes92a,hernquist93h} to produce remnants that were reasonable
matches to ellipticals; i.e., pure spheroid galaxies.
As merger simulations have become more sophisticated, the
specific details of the progenitor mass ratios \citep{naab99,naab03,cox08a},
and gas fractions \citep{cox06a,naab06d} are now understood to
critically shape the kinematic and photometric structure of merger remnants,
providing realistic merger scenarios for the formation of both elliptical
galaxy families found in nature 
\citep{kormendy96,emsellem07}:
low-mass disky fast rotators and high-mass boxy slow rotators.
Additional simulations demonstrate that mergers can also be responsible for 
early-type spirals and S0's under certain conditions; e.g., unequal-mass
spiral-spiral major mergers \citep{bekki98e}. 
If gas accretion continues after the spheroid formation a disk component may
reform \citep{barnes02,steinmetz02,abadi03b,governato07,governato09},
or even survive a major merger 
\citep{springel05b,robertson06c,hopkins09a}.
Furthermore, simulations predict that minor merging can play an
important role in spheroid and bulge growth 
\citep{bournaud07,naab07}, especially towards explaining the significant
size evolution of massive ETGs \citep{naab09b,hopkins10b,oser12}.
Besides merger-driven ETG formation,
the morphology-density relation \citep{dressler80a,postman84,dressler97}
has motivated environmental mechanisms for the transformation of disks
into ETGs, especially S0 galaxies in clusters, 
such as stripping and gas consumption
by SF \citep{larson80,bekki02c}, harassment \citep{farouki81,moore99},
and ram-pressure stripping \citep{quilis00}, but others argue that
the latter process is insufficient \citep[e.g.,][]{farouki80,boselli06}.
Another commonly invoked morphological transformation is
`preprocessing' by environmental processes in groups
before entering dense clusters \citep{zabludoff98,boselli06,wilman09,bekki11c}.
Finally, the secular growth of `pseudobulges' 
\citep{courteau96a,norman96,macarthur03,kormendy04,debattista06}
provides another channel for the morphological evolution of late-type disk
galaxies.

The diversity of galaxy transformational processes that can play a role
in blue-to-red migration is daunting. 
Attempts to disentangle the various quenching channels and ETG assembly modes
in the present-day Universe
\citep{mcintosh04,schawinski07b,kaviraj09c,gyory10,schawinski10a,greene12}
demonstrate the challenge in trying to improve our understanding of
the physics of red-sequence growth. 
For example, the so-called `green valley' between the blue cloud
and red sequence may include a significant fraction of quenched galaxies
migrating redwards \citep{martin07,mendel13}. Yet, there is a
rich variety of green-valley morphologies 
\citep{schawinski10a},
which can be produced in a number of ways.
Moreover, recent studies show that a significant portion of the local
ETG population has low-level, residual SF
\citep{kaviraj07c,schawinski07a,schawinski09c}, which extends and complicates
their redward migratory path.
For these reasons, our approach herein is to focus narrowly on the elliptical
(pure spheroid) subset of ETGs, and to isolate recently quenched examples;
i.e., strong {\it candidates}
for new, spiral-spiral merger-formed or `first-generation' ellipticals.
By attempting to catch new ellipticals transitioning to the red sequence,
we can better constrain
the role major merging plays in blue-to-red migration.
Furthermore, by limiting our investigation to {\it one}
morphological transformation channel (disk-disk major merging),
we can study the associated quenching processes and
test the modern merger hypothesis for the gas-rich\footnote{The
`gas-rich' distinction herein refers to dissipational or 
spiral-spiral merging, as opposed to gas-poor or dry.}
merger production
of red ellipticals \citep{springel05a,hopkins06b,hopkins08b}.

Trying to identify a connection between mergers and old red ellipticals
has proven elusive since \citet{toomre72} first suggested the link.
Identifying unambiguous first-generation ellipticals in the `gap'
between clearly merging systems and ancient ellipticals 
has considerable inherent
scatter owing to the sensitivity of each remnant's evolution to the properties
of the initial interaction and post-merger SF \citep{gyory10}.
Important attempts to correlate tidally-induced substructure and asymmetry
with colors and stellar populations have shown general agreement with the
merger hypothesis \citep{schweizer92,tal09,gyory10}. But others are finding that
the bulk of morphologically-disturbed red ellipticals with low-level recent SF are
either the product of major dry mergers or minor mergers \citep{kaviraj10b}.
Other studies have attempted to identify young elliptical galaxies; e.g.,
star-bursting ultraluminous infrared galaxies
\citep[ULIRGs][]{sanders88a,genzel01,dasyra06b} and
very recent post-merger remnants with strong morphological disturbances
\citep{hibbard96,rothberg04,carpineti12}
are rare examples cited as ellipticals in formation.
Additionally, post-starburst (E$+$A or K$+$A) galaxies 
with strong Balmer (A-star) absorption but
no emission from ongoing SF \citep{quintero04,goto05}
are clear examples of a recently quenched population that is tied
to galaxy merging because of the high incidence of morphological disturbance
\citep{zabludoff96,blake04,yang04,bekki05,goto05,yang08}, but 
it is unclear whether this phenomenon accompanies
all gas-rich mergers or only a special subset.
Local Universe post-starburst galaxies are found
preferentially at low masses 
\citep[$<10^{10}~{\rm M}_{\sun}$][]{wong12}, and at a much
lower frequency \citep[$\sim 0.1\%$][]{goto07b,wong12} 
compared to major spiral-spiral mergers
\citep[$\sim1\%$][; McIntosh et al., in prep.]{darg10a}.
Under the assumption that all such galaxies were the result of a recent
gaseous merger, 
this difference may be further exacerbated depending on whether the
E$+$A phase lasts $\sim1$\,Gyr (typical A-star lifetime) or 
$\leq 0.1-0.3$\,Gyr \citep{snyder11} compared to the
visibility time-scales for first close passage
and final coalescence in interactions and mergers, where the strong
morphological asymmetries are identifiable
\citep[$<0.5$\,Gyr,][]{lotz08b,lotz10a,lotz10b}.
Therefore, we use an alternative approach to identify plausible
first-generation ellipticals under the assumption that young gas-rich merger
remnants {\it should} experience a relatively brief blue
phase whether the merger fueled a strong starburst or a modest SF enhancement.
Then we adopt a method similar to
\citet{whitaker12a} to identify the recently quenched subset of blue 
ellipticals, which are spectroscopically quiescent and have 
unusually young luminosity-weighted stellar ages.

A final complication in identifying and tracking first-generation 
ellipticals migrating
to the red sequence is the rejuvenation of previously passive
ellipticals that experienced a temporary minor SF event \citep{thomas10}.
In our standard hierarchical cosmological model, the stellar build up of
galaxies over time is a complex process involving both the accretion of gas
and the addition of already-formed stars through galaxy merging 
\citep[e.g.,][]{oser10}. 
When a quiescent ETG experiences a gas-rich minor merger, or
otherwise accretes star-forming gas, a 
``frosting'' of low-level SF \citep{trager00a}
may add only $\sim1\%$ to its mass, but have enough
hot O and B stars to produce a brief global blue color owing to the
extended nature of the SF.
Evidence of young stars and low-level recent SF has been 
observed in low-redshift ETGs \citep{yi05,kaviraj07c,schawinski07a}, 
including those
with disks \citep{shapiro10,fang12,salim12a} and pure elliptical subsets
\citep{sanchezblazquez09,zhu10}.
The accretion of small companions
builds up the mass of existing red ellipticals and is thought to play
a significant role in their rapid size evolution within $0<z<2$
\citep{vanderwel08,bezanson09}.
Many moderately blue (i.e., green-valley)
ETGs, even those with asymmetric morphologies, may not be newly formed
spheroidal galaxies. 
Moreover, high-resolution simulations show that the tidal debris seen around
massive ellipticals can be produced by major and minor merging alike
\citep{feldmann08}.
In other words, the mass assembly of ellipticals is not simply a one-way 
channel from blue to red via gas-rich major merging. 
By focusing on recently quenched ellipticals (RQEs), 
we are constraining our study to merger-built
galaxies migrating redwards.  These are plausibly first-generation 
ellipticals, but
they are also consistent with `second-generation' examples; i.e., a past
merger remnant that was recently rejuvenated and subsequently quenched.
We caution that unambiguously
distinguishing new ellipticals (with central concentrations of new stars) from
frosted ellipticals (with extended young stellar populations)
requires spatially resolved star formation histories (SFHs) 
from IFU spectroscopy
and is beyond the scope of this study.
A more detailed study of the SFHs of blue ellipticals will be presented
in a subsequent paper (Haines et al., in prep.).

In this paper, we analyze a sample of 1500 visually selected
elliptical (pure-spheroid) galaxies with unusually blue optical colors.
These objects are drawn from a large and complete selection of 
massive (${\rm M}_{{\rm gal,}{\star}}\geq10^{10}~h^{-2}~{\rm M}_{\sun}$),
centrally-concentrated galaxies with $0.01<z\leq 0.08$ 
in the Sloan Digital Sky Survey \citep[SDSS;][]{york00}.
In \S~\ref{sec:Analysis}, 
we study the structure, spectroscopic emission and rest-frame colors
of unusually blue ellipticals. This analysis allows us to identify
a robust subset of blue ellipticals that are non-star-forming.
In \S~\ref{sec:Quenched}, we isolate and characterize a new population of 
recently quenched systems among the non-star-forming
blue ellipticals, and we discuss their plausible
quenching mechanisms. These objects represent an objectively selected and
statistically useful sample of first-generation elliptical candidates
for further study.
Throughout this paper we calculate comoving distances in the $\Lambda$
cold dark matter ($\Lambda$CDM)
concordance cosmology with $\Omega_{\rm m} = 0.3$,
$\Omega_{\Lambda} = 0.7$, and assume a Hubble constant of
$H_0 = 100\, h $\,km\,s$^{-1}$\,Mpc$^{-1}$. All SDSS magnitudes are on the
AB system such that $m_{\rm AB}=m+\Delta m$, where
$\Delta m=(-0.036,+0.012,+0.010,+0.028,+0.040)$ for
$(u,g,r,i,z)$ \citep{yang07}.

\section{Selection of Blue Elliptical Galaxies}
\label{sec:sample}
As motivated in the Introduction, the primary goal of this study is to
identify and analyze 
high-mass elliptical (pure spheroid) galaxies at low redshift
that are plausibly transitioning redwards.
To isolate a statistical sample of such galaxies that is mass-limited,
volume-limited and has high spectroscopic completeness 
we employ the SDSS Main Galaxy Sample \citep[MGS,][]{strauss02}
and apply the following 3-step selection:
(1) isolate a large sample of 
blue-cloud galaxies with redshifts $z\leq0.08$ 
and stellar masses bracketing the bimodal mass
scale of $3\times 10^{10}~{\rm M}_{\sun}$,
(2) apply an automated high-concentration cut to
identify the bulge-dominated subset (herefter blue ETGs), 
and (3) visually distinguish elliptical galaxies from other
more dominant morphologies found in the blue ETG population
(e.g., spiral and disk galaxies with prominent bulges).

It is important to point out that employing a simple high-concentration cut
to automatically extract ETGs from the high-mass blue cloud includes a large
mix of {\it non-elliptical} morphologies. This fact motivates the
additional visual selection criteria which provides a robust separation
of elliptical galaxies from the larger population of blue ETGs.
Our selection identifies the 1500 most optically blue ellipticals 
among a complete population of over 60,000 high-mass, low-redshift galaxies in
the SDSS fourth data release \citep[DR4,][]{adelman06}. These objects
make up only 2\% of high-mass galaxies, but are potentially
evolutionarily quite important as they are strong candidates for new
merger-formed or 'first-generation' ellipticals.
Here, we discuss the details of our blue elliptical galaxy
selection process.

\subsection{Stellar Masses and Optical Colors}
\label{sec:bluegxs}
We use the New
York University Value-Added Galaxy Catalog \citep[NYU-VAGC,][]{blanton05}
reprocessing of the MGS for spectroscopic target selection, which is
limited to all extended sources with $r<17.77$ magnitudes.
We calculate stellar masses following \citet{bell03b}:
\begin{equation}
\begin{split}
\log_{10}({\rm M}_{{\rm gal,}{\star}}/h^{-2}{\rm M}_{\sun}) = -0.406 + 1.097[^{0.0}(g-r)] \\ -0.4(^{0.0}M_r - 5\log_{10}h-4.64) .
\end{split}
\end{equation}
This relation adopts a \citet{kroupa01} IMF.
The $^{0.0}(g-r)$ color and absolute $r$-band magnitude $^{0.0}M_r$
have been $K$-corrected \citep{blanton07} and evolution-corrected 
for simple passive luminosity fading \citep{blanton03c} to redshift $z=0$.
The NYU-VAGC provides significantly improved photometry for all SDSS galaxies.
We use SDSS Petrosian magnitudes and colors 
corrected for Galactic extinction \citep{schlegel98}.
For galaxies with early-type morphologies, defined by an $r$-band
central-light concentration cut $c_r=R_{90}/R_{50}\geq2.6$
\citep[using the radii containing 90\% and 50\% of the Petrosian flux, see;
e.g.,][]{strateva01}, we correct $^{0.0}M_r$ by
$-0.1$\,mag for well-known missing flux \citep{blanton03c}.
The Bell et al. stellar masses have 20\% random uncertainties 
and 0.10--0.15 dex
systematic error from a combination of effects including dust, stellar
population ages and bursts of star formation.
Tests based on the same stellar mass estimates 
show that the SDSS is very complete out to $z=0.06$ for
${\rm M}_{{\rm gal,}{\star}}\geq10^{10}~h^{-2}~{\rm M}_{\sun}$
\citep[][see Appendix A]{vandenbosch08}, and remains fairly complete to
$z\leq0.08$ for
${\rm M}_{{\rm gal,}{\star}}\geq 2\times 10^{10}~h^{-2}~{\rm M}_{\sun}$.
Selections of $0.01<z\leq 0.08$ and 
${\rm M}_{{\rm gal,}{\star}}\geq10^{10}~h^{-2}~{\rm M}_{\sun}$
yields 63,454 DR4 galaxies.

\begin{figure*}
\center{\includegraphics[scale=0.63, angle=0]{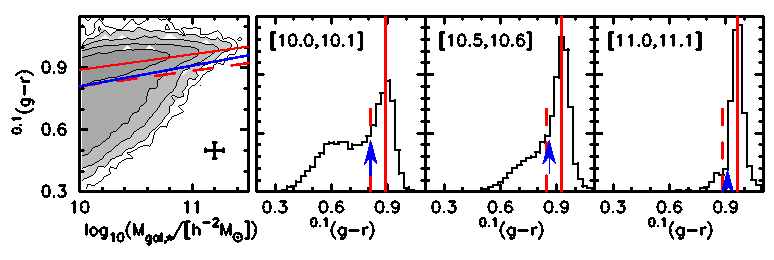}}
\caption[]{
Massive blue galaxy sample selection.
{\it Left:}
Contour plot of the color-stellar mass distribution for the full sample
of 63,454 SDSS galaxies with $0.01<z<0.08$. The error bar shows the typical
uncertainties.
{\it Right panels:}
The color distributions for three mass
$\Delta (\log_{10}{\rm M}_{{\rm gal,}{\star}})=0.1$ bins.
The centroid of the red sequence is shown with red lines in each panel,
a blueward vertical shift by $2\sigma_{(g-r)}=0.08$ is depicted with
red dashed lines.
The blue line and vertical arrows indicate the empirical cut that we employ
and is defined by the color at which the galaxy
population deviates from the roughly log-normal color distribution
of the red sequence. Colors are based on
Petrosian magnitudes corrected for Galactic extinction and
$K+$evolution corrected to $z=0.1$. Stellar mass estimates are based
on $(g-r)$ color and \citet{bell03b} M/L ratios.
\label{fig:blueseln}}
\end{figure*}

We employ an empirically defined color cut to distinguish blue-cloud 
and red-sequence galaxies.
The left-most panel of Figure~\ref{fig:blueseln}
shows the stellar mass and $(g-r)$ color space for the full sample of
63,454 high-mass, low-redshift galaxies. The colors are 
$K+$evolution corrected to $z=0.1$ to more closely match the median redshift
of our selection ($z_{\rm med}=0.066$).
The right panels show the distributions in color for 
three fixed mass slices
of width $\Delta (\log_{10}{\rm M}_{{\rm gal,}{\star}})=0.1$.
The well-known bimodality in galaxy colors \citep[e.g.,][]{baldry04}
remains apparent at 
$\sim 10^{10}~h^{-2}~{\rm M}_{\sun}$, but disappears as the blue population
drops off sharply at higher masses. The red sequence is clear at all masses
and we use this feature to define blue/red cuts.
The red-sequence centroid (solid red lines) is given by
\begin{equation}
^{0.1}(g-r) =\, ^{0.1}A + m[\log_{10}({\rm M}_{{\rm gal,}{\star}}/h^{-2}{\rm M}_{\sun})-10.0] ,
\end{equation}
with $^{0.1}A=0.89$ and slope $m=0.075$. The dashed red lines
($^{0.1}A=0.81$, $m=0.075$) show a common red/blue division defined by
a blueward shift from the red sequence by some factor, in this case twice
the typical color error (0.04 mag). 
In this paper, we adopt an {\it empirically-derived} blue/red cut
($^{0.1}A=0.81$, $m=0.1$, blue line and arrows) which is defined by the color
at which the galaxy population deviates bluewards from the roughly log-normal
color distribution for red galaxies. The empirical cut matches the common
division at low mass but for more massive galaxies it includes a larger
number of galaxies into the blue sample. Our
empirical blue color selection includes 38.9\% (24,659 galaxies)
of the high-mass population at $0.01<z\leq0.08$.

\subsection{Morphologies}
\label{sec:blueETGmorphs}

To isolate a robust sample of unusually 
blue ellipticals from the large sample of high-mass blue galaxies selected
in \S~\ref{sec:bluegxs}, we take a hybrid approach
of {\it pre-selecting} all ETGs based on a popular and
automatic high-concentration criterion, 
followed by visual classification to remove 
bulge-dominated disk galaxies (early-type spirals and S0's) and other
contaminants.
This approach has been used successfully by others with similar data and
objectives \citep[e.g.,][]{zhu10}.
In what follows, we include a summary
of the full richness of the visual morphology inventory 
of low-redshift, high-mass galaxies that
are blue and centrally concentrated. Additionally, we describe tests 
demonstrating the robustness of our visual classifications.

\subsubsection{Central Concentration}
\label{sec:ConcCut}

For ETG preselection, we apply the common SDSS 
$r$-band concentration cut $c_r=R_{90}/R_{50}\geq2.6$ to
automatically separate high-concentration (bulge-dominated) galaxies
\citep[e.g., see][]{strateva01,hogg02,bell03b,kauffmann03b} 
from the bulk of the blue population which is
disk-dominated with low central light concentrations
\citep[e.g.,][]{blanton09}. 
We note that all galaxies are resolved well enough ($R_{50}>3$\,pixels)
to apply this simple concentration criterion.
We find that one-third of high-mass blue galaxies have $c_r\geq2.6$.
It is important to point out that
naively applying a pure concentration cut to the full sample of 
high-mass blue galaxies results in an unrealistically large portion (13.3\%)
of crudely defined blue ETGs. From the analysis of Blanton \&
Moustakas, we expect that a large portion of these blue ETGs correspond
to luminous early-type spirals (Hubble type Sa) and non-red-sequence S0
galaxies in contrast to the dominant E/S0 morphologies found typically
among high-mass red ETGs. 
We confirm the expected high content of bulge-dominated spirals and disks
among automatically selected blue ETGs using visual inspection in 
\S~\ref{sec:BCmorphs}.

\subsubsection{Visual Classification Scheme}
\label{sec:VisScheme}

We use visual inspection to extract the {\it subset} of blue ellipticals 
(i.e., remove bulge-dominated disk galaxies)
from the full set of 8421 high-mass blue ETGs selected in \S~\ref{sec:ConcCut}.
Visual inspection has the added benefit of identifying
{\it asymmetric or peculiar} elliptical galaxies and 
highly-disturbed, recent spheroidal post-merger candidates.
In particular, merger simulations predict short-lived tidal features
(e.g., tails, loops, shells, plumes) associated with young spheroidal
remnants.
Despite its subjective nature, visual classification maintains
an advantage over automated classification schemes 
\citep[e.g.,][]{conselice03a,lotz08b} in the ability to distinguish especially
lower-surface brightness
peculiarities and asymmetries associated with recent merging activity
from the rich variety of structures found in normal spiral galaxies. 
Our classification scheme consists of the following morphological types:
\begin{itemize}
  \item {\bf S} $=$ {\it Spirals}
exhibit one or more {\it clear} disk resonance feature (e.g.,
spiral arms, stellar bar, or inner ring), or are highly flattened with a
central dustlane.
  \item {\bf iD} $=$ {\it Inclined disks} are elongated galaxies
with disk-like outer isophotes but unclear spiral signatures.
  \item {\bf E} $=$ {\it Elliptical} or spheroidal galaxies with bright centers
and smooth light profiles showing
little or no asymmetric features.
  \item {\bf pE} $=$ {\it Peculiar elliptical} galaxies with one or more of the
following morphological disturbances
consistent with recent tidal activity:
excess outer light, asymmetric outer
isophotes, shells, asymmetric dustlane, blue core, or clearly 
dust-reddened core.
  \item {\bf SPM} $=$ Highly-disturbed, {\it spheroidal post-merger}
galaxies with
strong global asymmetries suggesting a recent merger origin. 
  \item {\bf U} $=$ Galaxies with {\it uncertain} morphologies: these 
objects often appear round or elliptical with
subtle ({\it unclear}) structural features suggestive of
face-on spirals or spheroids, but too faint to detect clearly. 
\end{itemize}

\subsubsection{Inventory of Blue-cloud High-concentration Types}
\label{sec:BCmorphs}
Each galaxy was independently inspected by the lead author plus
three students (AMC, TH, \& JM) and assigned one of the types from our
classification scheme.
The classifications were based on inspection of $gri$-combined color
postage stamps with fixed sensitivity scaling available
from the SDSS Image List Tool\footnote{
{\texttt http://cas.sdss.org/astro/en/tools/chart/list.asp}.}.
We note that the use of fixed scale image cutouts for visual classification
of bright galaxies is the standard employed in the largest efforts at
both low and high redshift 
\citep[e.g.,][]{lintott11,kartaltepe14}.
We removed 356 galaxies that have evidence of an ongoing
tidal interaction with
a close companion. These galaxies could be a recent merger remnant but are
undergoing another interaction making interpretation of recent past
activity difficult.
Finally, we identified and omitted a small fraction (1.9\%) of
galaxies with questionable morphologies and/or colors
owing to severe scattered light contamination from nearby bright stars,
and we excluded 12 (0.1\%) objects that are
image artifacts (e.g., satellite trails, stellar diffraction spikes).
After these exclusions, our final sample contains 7890 high-mass blue ETGs
with careful visual classifications.

\begin{figure*}
\center{\includegraphics[scale=0.75, angle=0]{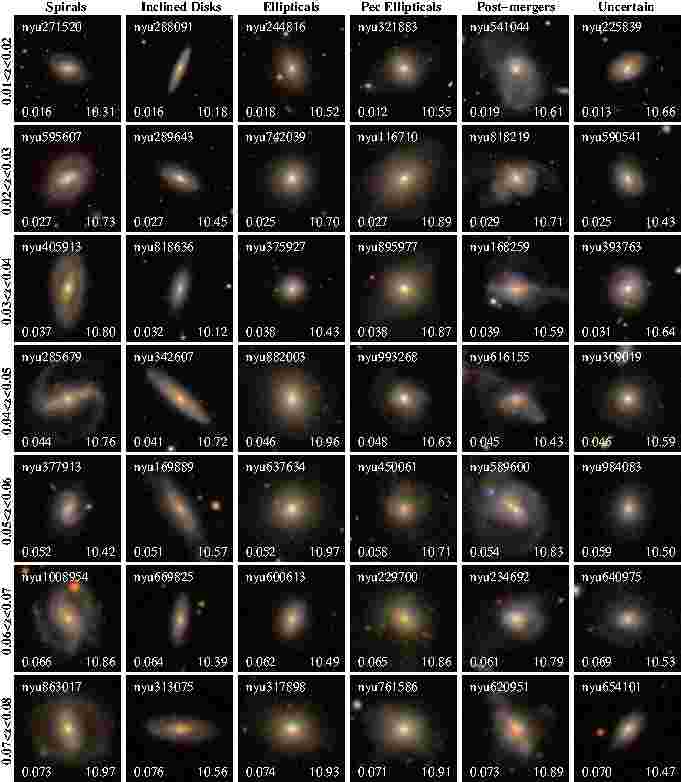}}
\caption[]{
Examples of the six visual classification types from left to right:
spirals (S), inclined disks (iD), ellipticals (E), peculiar ellipticals
(pE), spheroidal post-mergers (SPM), and galaxies with uncertain morphology (U).
Each column contains postage stamp images of galaxies from
seven $\Delta z=0.01$ redshift bins spanning $0.01<z<0.08$.
Examples are selected from the subsets with
{\it high} classifier agreement (minimum three out of four).
All images are $40\times40$ kpc
cutouts of $gri$-combined color images with fixed sensitivity scaling
downloaded from the SDSS Image List Tool, and include the galaxy
identification number \citep[from the DR4 NYU-VAGC,][]{blanton05},
redshift and log stellar mass in units of
$\log_{10}(h^{-2}{\rm M}_{\sun})$.
\label{fig:examples}}
\end{figure*}

We define {\it high (classifier-to-classifier) agreement} to be a
minimum of three out of four classifiers in agreement. We find this level
of agreement for 86.7\% of the high-mass blue ETG sample.
We tabulate the visual classification
summary in Table~\ref{tab:classifs}.
Our classification scheme divides the blue ETG sample into three basic subsets:
(i) 
galaxies with {\it clear} spiral/disk morphologies (S, iD types) which,
by definition,
cannot be remnants of recent {\it disk-destroying} major mergers\footnote{High-mass, low-redshift disk galaxies have relatively low gas fractions, as
such we do not expect disk regrowth following a major merger as seen
for extreme gas fractions \citep[e.g.,][]{springel05b}.};
(ii) 
plausible merger remnants (E, pE and SPM types)
spanning a range of post-merger characteristics;
and (iii) 
galaxies that lack definitive morphological structure or features
in imaging at the depth of the SDSS (U type). In Figure \ref{fig:examples},
we show illustrative examples of our classification types from eight
redshift bins spanning $0.01<z<0.08$.
Clearly, high-mass blue ETGs
in the local cosmological volume have a wide range of morphologies;
We reiterate that a crude concentration cut 
is poor for selecting a pure elliptical galaxy sample from the blue cloud;
i.e., more than 40\% of $c_r\geq 2.6$ blue galaxies have clear
spiral and/or disk (including S0) features (see Table~\ref{tab:classifs}).
This is an important result of this classification analysis.
In the next section, 
we outline a number of tests we perform to establish the validity
of our visual selection of smooth, pure-spheroid elliptical galaxies which
are the primary subject of this paper.

It is worth noting that the
visually-identified SPM, pE and E subsets of the high-mass, blue ETGs 
represent three morphological bins
that could plausibly span 
a {\it qualitative} time sequence since merging. At one extreme
we find galaxies that appear to be dynamically relaxed ellipticals
with little/no evidence of recent tidal activity, contrasted by those at
the other extreme that
appear to be freshly coalesced with very disturbed morphologies. The visual
appearance of the blue SPM, pE and E types (see Fig.~\ref{fig:examples})
qualitatively matches the evolution of tidal
features in merger simulations.
In concert with the expected
blue-to-red migration, simulated merger remnants evolve morphologically as
the post-merger tidal features (e.g., tails, loops, shells, plumes) quickly
fade around the dynamically relaxing young spheroidal galaxy
during the first $\sim1$ billion years after coalescence
\citep[e.g.,][]{barnes92a,barnes96}.
As such, we expect distinct stages of evolution between a major
spiral-spiral interaction and a red elliptical including
(i) a highly-disturbed, dynamically-young remnant,
(ii) a spheroid-dominated core with tidal features, and
(iii) a bluer-than-normal elliptical.
In this scenario, the low number of clear SPMs (110) and pEs (124),
compared to 1368 featureless blue Es, is consistent
with the rapid disappearance of strong tidal features in the first few
$10^8\,{\rm years}$ after coalescence as the remnant relaxes.
SPMs are perhaps the cleanest examples of {\it very young} new spheroidal
galaxies and, as such,
are an interesting population in their own right with regard to the
merger hypothesis. A detailed analysis of their properties is the 
subject of a forthcoming paper. Additionally, we have completed a study
of the radial SFHs of a small representative sample of these blue
E, pE and SPMs (Haines et al., in prep.).

\subsubsection{Classification Process and Tests}
\label{sec:classTests}

To achieve meaningful results based on visual identification of galaxy subsets
requires that we understand the robustness of our eyeball classifications to
meet two important goals: (1) distinguish ellipticals from spirals; and
(2) identify evidence of recent tidal activity. Besides the
inherent subjectivity of human classifiers, changes in
image resolution and sensitivity between galaxies of different brightnesses
over a range of distances will impact the quality of visual classifications.
To address these issues, we 
perform a number of tests to check the validity of our classifications.
Additionally, we establish in \S~\ref{sec:Struct} that our sample of visually
identified blue ellipticals are well matched to red ETGs (typically E/S0) 
in terms of quantitative structural parameters.

The most critical test of our classifications is possible by examining data
from the SDSS Stripe82. As described in
Data Release 7 \citep{abazajian09}, this $>250$\, deg$^2$ region extending
from $-50\degr<\alpha<59\degr$ along the celestial
equator in the Southern Galactic Cap was imaged 20--40 times.
Thus, Stripe82 provides imaging for $\sim5\%$ of our blue ETG selection
that is $\sim2$\,mag deeper than the standard SDSS images. We find no 
significant changes in our visual morphologies based on examination of
the Stripe82 subset. In particular, 95\% (69/73) of high-agreement elliptical
classifications remain E or pE types based on Stripe82 data. Three-quarters
of the remainder may be S0 galaxies while only one
showed a clear disk component. Most importantly, all RQEs
(see \S~\ref{sec:Quenched}) with Stripe82 data are
confirmed morphologically by the deeper imaging.

Additionally, we test how our elliptical classifications depend on apparent
brightness and angular size.
First, we find that the fraction of high-agreement E$+$pE types among high-mass
blue ETGs remains fairly constant in three magnitude bins: 
24\% ($r<15$\, mag), 20\% ($15<r<16$\,mag) and 19\% ($r>16$\,mag). The slight
decrease with increasing magnitude is the result of greater numbers of
uncertain (U-type) morphologies at lower stellar masses and higher redshifts
as discussed in the next paragraph.
Secondly, we find that the level of elliptical E or pE agreement between
classifiers remains unchanged for angular half-light sizes
$R_{50}<2$, $2<R_{50}<3$ and $R_{50}>3$ arcseconds.

Additionally,
we thoroughly test our classifications dependence on redshift $z$ and galaxy
mass ${\rm M}_{{\rm gal,}{\star}}$ and find the following: 
(i) no decrease in the level of classifier
agreement for identifying S/iD types over all ${\rm M}_{{\rm gal,}{\star}}$
and $z$;
(ii) 80\% of iD classifications have an axial ratio of $b/a<0.5$;
(iii) agreement on E classifications is independent of $z$ but depends on
${\rm M}_{{\rm gal,}{\star}}$ such that lower-mass galaxies have lower
agreement, which is expected given the fact that lower-mass ellipticals
tend to be more disky \citep[e.g.,][]{kormendy96} and may be classified 
as iD instead; and
(iv) the majority of blue ETGs have a {\it clear} (non-U) classification
except those with 
${\rm M}_{{\rm gal,}{\star}}\leq 3\times 10^{10}~h^{-2}~{\rm M}_{\sun}$ 
and $z\geq 0.06$, of which one-third are high-agreement U types.

While the number of pE and SPM types is too small to quantify conclusive
trends, we note that we detect SPMs at all $z$ and
${\rm M}_{{\rm gal,}{\star}}$ 
(98\% below $10^{11}~h^{-2}~{\rm M}_{\sun}$), and we find pEs at
all redshifts but 95\% 
have ${\rm M}_{{\rm gal,}{\star}}>2\times 10^{10}~h^{-2}~{\rm M}_{\sun}$ 
with no obvious $z$ bias among
the few lower-mass identifications.
We estimate our ability to identify morphological disturbances
by looking at our spiral feature
detection. We expect that
if resolved well enough, a given sample of spiral disk galaxies
with a typical distribution of axial ratios from edge-on to face-on
will have a majority with clearly identifiable spiral features -- our S
definition. At fixed angular resolution, the physical resolution decreases 
from either smaller physical size 
\citep[lower-mass disks have smaller average size; e.g.,][]{shen03},
or smaller apparent size from increased distance. Therefore, 
we expect the visibility of spiral structure to be more sensitive to resolution
changes than the overall disky isophotes of inclined disks -- our iD type.
We test which galaxies 
with high S/iD agreement are dominated by S and by iD classifications and
we find that iD dominate over S identifications only for galaxies with
${\rm M}_{{\rm gal,}{\star}}\leq 3\times10^{10}~h^{-2}~{\rm M}_{\sun}$ and
$z\geq 0.05$, which is quite similar to the parameter space where U types
increase.

We conclude that our pE and SPM identifications are robust for
${\rm M}_{{\rm gal,}{\star}}>3\times10^{10}~h^{-2}~{\rm M}_{\sun}$ at
all redshifts, and down to $10^{10}~h^{-2}~{\rm M}_{\sun}$ for $z<0.05$,
but we cannot ascertain whether the lack of disturbed ellipticals
below $2\times10^{10}~h^{-2}~{\rm M}_{\sun}$
is real or the result of resolution bias. Additionally,
our visual classifications distinguish pure-spheroid
ellipticals from early-type spirals and bulge-dominated disks 
with high classifier-to-classifier 
agreement over most
masses and redshifts. 

\subsection{Refined Sample for Analysis}
\label{sec:BEsamp}

For the remainder of this study we will focus on the 1492 unusually
blue ellipticals
with high visual classification agreement as our final
sample for analysis. This sample includes 1368 (normal, E type) and
124 (peculiar, pE type) elliptical galaxies. We opt
to exclude SPMs from our analysis, which are not ellipticals (yet).
At $z\leq0.08$, these blue ellipticals make up only
3.7\% of all high-concentration galaxies with stellar masses between
$10^{10}$ and $10^{11}~h^{-2}~{\rm M}_{\sun}$. This percentage is consistent
with the findings of \citet{kannappan09} based on several samples of 
very low-redshift galaxies.
Additionally, blue ellipticals make up a
tiny fraction of the 25,000 most-massive blue galaxies in this SDSS DR4
volume -- most of which have low concentrations
($c_r<2.6$) consistent with disk-dominated systems.

The blue elliptical sample provides the best candidates to test the modern 
major merger
hypothesis for the formation of new elliptical galaxies. We expect that such
galaxies include older (pre-existing) ellipticals that are modestly blue
as the result of non-major-merger assembly processes.
For example, the peculiar
ellipticals exhibit a host of tidal features including loops, tails and
distorted dust features that can arise from either major or minor mergers 
between
existing ellipticals and cold-gas disks \citep{feldmann08}. 
In what follows, we use a suite of SDSS-derived quantities to confirm the
spheroid nature of these visually identified ellipticals and to
divide them into subpopulations of interest for closer scrutiny.

\begin{table}
\caption{Visual classification summary.}
\label{tab:classifs}
\begin{tabular}{lccccc}
\hline\hline
Classification & $N_{\geq 3}$ & $N_4$ \\
\hline
Elliptical (E)              & 1368 & 1030 \\
Peculiar elliptical (pE)      &  124 &   52 \\
Spheroidal post-merger (SPM) &  110 &   43 \\
Spiral/inclined disk (S/iD)$^{\rm a}$ & 2834 & 2236 \\
Uncertain (U)               & 2408 & 1267 \\
\hline
Total (of 7890)             & 6844 & 4628 \\
\hline
\end{tabular}
\vskip 8pt
\begin{minipage}{\hssize}
Total number of high-concentration ($c_r\geq2.6$), blue-cloud galaxies
that are
classified as the same type
by at least three classifiers $N_{\geq 3}$
and by all four classifiers $N_4$.\\
$^{\rm a}$ Agreement is defined as any combination of S and iD classifications.
\end{minipage}
\end{table}

\section{Properties of Visually-Selected Blue Elliptical Galaxies}
\label{sec:Analysis}

In \S 2, we visually identified a stellar mass-limited sample of blue 
elliptical galaxies with normal (E) and peculiar (pE) morphologies.
In this section, we compare the structure of these galaxies against 
control samples of automatically-selected red-sequence ETGs
and blue-cloud late-type galaxies (LTGs). Next, we divide the blue
ellipticals (E$+$pE types) into different
emission types using several spectroscopic diagnostics. Lastly, we
explore the color versus velocity dispersion distributions of our sample
split by emission type and compare these to the underlying galaxy population.

\subsection{Structure}
\label{sec:Struct}

\begin{figure*}
  \center{
  \includegraphics[scale=0.4, angle=0]{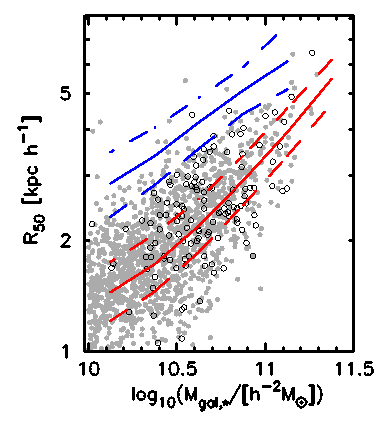}
  \includegraphics[scale=0.4, angle=0]{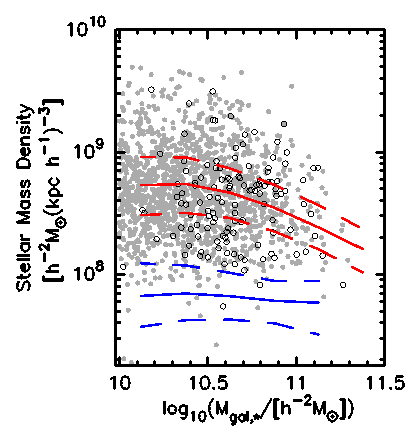}
  \includegraphics[scale=0.4, angle=0]{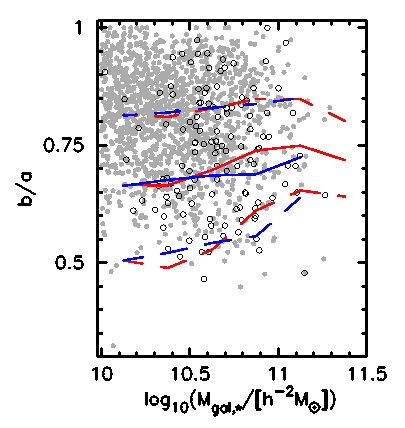}
  }
\caption[]{Stellar mass versus half-light size ({\it left}), half-light
stellar density ({\it center}), and axial ratio ({\it right}) of blue elliptical
galaxies with {\it high} classifier-to-classifier agreement. 
In each panel, the structure of visually-identified
(normal) ellipticals (Es, grey circles) and
peculiar ellipticals (pEs, black open circles) are shown
relative to the median (solid line) plus 25\%- and 75\%-tiles (dashed lines)
of the
distributions of two control samples: 16,241 blue LTGs ($c_r<2.6$) in blue,
and 32,349 red ETGs ($c_r\geq2.6$) in red. The control sample statistics are
calculated in bins of
$\log_{10}({\rm M}_{{\rm gal,}{\star}}/h^{-2}{\rm M}_{\sun})=0.25$ and limited
to those containing $N_{\rm bin}>10$ galaxies.
Half-light size is based on $r$-band Petrosian fluxes. The stellar density
is given by
$\frac{1}{2}{\rm M}_{{\rm gal,}{\star}}/(\frac{4}{3} \pi R_{50}^3)$ ({\it right}). The axial ratios are $r$-band isophotal.
\label{fig:Struct}}
\end{figure*}

To establish the fidelity of our visual selection of blue ellipticals,
we compare their structure against that from two control samples in
Figure~\ref{fig:Struct}. In the left panel, we plot the stellar masses and
half-light sizes $R_{50}$ of blue elliptical galaxies, and 
show the median, 25\%-tile and 75\%-tile empirical mass-size relations for 
32,349 red ETGs ($c_r\geq 2.6$) and 
16,241 blue disk-dominated LTGs ($c_r<2.6$).
The sizes are based on the radius (petroR50\_r) of the circular
aperture containing 50\% of the $r$-band Petrosian flux \citep{strauss02}.
The unique mass-size relations for disk and spheroid-dominated galaxies are
well known \citep[e.g.,][]{shen03}.
At a given stellar mass, the blue ellipticals have size distributions
that nearly match that of red ETGs and are quite different from the typical
sizes of blue LTGs. Put another way, the mass-size relations of blue
Es and pEs are clearly {\it inconsistent} with the relation for
disk-dominated galaxies from the blue cloud.
Likewise, as shown in the center panel, our galaxies are clearly more
dense than blue disks and have similar densities as red ETGs.

While the mass-size and mass-density 
relations of blue ellipticals are similar to
spheroid-dominated 
galaxies on the red sequence, there are a few minor differences
in Figure~\ref{fig:Struct}
worth noting.  First, blue Es are offset to slightly larger sizes
than red ETGs and, on average, blue pEs are somewhat less compact
and dense than their normal counterparts. Slightly larger sizes could
be the result of new stars added preferentially at larger radii.
Moreover, the pE subset have minor tidal features which likely
result in additional outer light and larger $R_{50}$ values. We note
that very few pEs are identified below 
${\rm M}_{{\rm gal,}{\star}}=2.5\times 10^{10}~h^{-2}~{\rm M}_{\sun}$,
but whether this is real or a selection effect remains inconclusive
(see \S~\ref{sec:classTests};
we discuss this point in more detail in \S~\ref{sec:colors}).
Secondly, blue ellipticals tend to have moderate stellar masses
with very few found above 
${\rm M}_{{\rm gal,}{\star}}=10^{11}~h^{-2}~{\rm M}_{\sun}$; i.e.,
any that are recent end products of major gas-rich merging will
become only moderate-mass red ellipticals. This is consistent with the idea
that dry merging, whether minor or major, dominates the mass assembly of
the largest ellipticals. 

We plot the axial ratios $b/a$ as a function of stellar mass for our sample
in the right panel of Figure~\ref{fig:Struct}. The axial ratios are based
on the isophotal semi-major and semi-minor axes (isoA\_r, isoB\_r)
measured from the $r$-band images. 
The two control samples both span roughly the same $b/a$ range
at a given stellar mass, with a more pronounced trend toward round
(high $b/a$) red ETGs above $10^{11}~h^{-2}~{\rm M}_{\sun}$ 
as found in a similar analysis
on massive quiescent galaxies by \citet{vanderwel09b}. 
The blue LTG values are counter-intuitive on first inspection because one
expects a random distribution of pure disks to extend to lower $b/a$ values
from edge-on and highly-inclined systems. But, highly-inclined disk galaxies 
tend to have higher concentrations and redder colors \citep{maller09},
which are both excluded by our blue LTG selection. Thus, the blue LTG control
sample is skewed toward rounder, less-inclined and face-on, galaxies.

The main result from the axial ratio distribution is that visually-selected
blue ellipticals are much rounder on average
than red ETGs over the full range of masses we study. This is not
surprising given the fact that the red ETG sample contains S0 and early-type
spiral galaxies as well as ellipticals. As such, blue ellipticals make up
only a fraction of red ETG progenitors (e.g., they will not evolve into red
S0's). 
More than half have $b/a>0.8$ which is roughly the upper quartile
for red ETGs.
The small subset of blue pEs have more elongated axial ratios than blue Es
of similar stellar mass. We interpret the difference
between blue E and pE axial ratio distributions to be the impact of
asymmetric and irregular tidal features on the $b/a$ measurement.
The axial ratios of our visually-selected ellipticals (E$+$pE) demonstrate the 
fidelity of our visual classifications to exclude S0's.
For comparison, 95\% of our blue ellipticals have $b/a>0.6$, a cut that
\citet{zhu10} employed to select a high-fidelity sample of low-redshift 
elliptical galaxies from the SDSS.
\citet{vanderwel09b} point out that the only mechanism to produce very round
spheroidal galaxies is major merging. In what follows, we will examine
these galaxies more closely to attempt to shed light on
whether they are newly-formed, first-generation ellipticals 
or previously-formed spheroids with a population of young stars.

\subsection{Spectroscopic Types}
\label{sec:EmTypes}

The SDSS fiber spectroscopy provides a number of useful diagnostics
for understanding the physical nature of blue ellipticals.
The $3\arcsec$ diameter
fibers correspond to a $0.4-3.2~h^{-1}~{\rm kpc}$ at the redshifts of
our sample and encompasses roughly 15--50\% of the stellar mass
of each galaxy.
We make use of the MPA-JHU emission-line analysis of the SDSS DR7 spectra
\citep[for details, see][]{kauffmann03c,brinchmann04} which provides
stellar continuum-subtracted [based on an updated version of the
\citet{bruzual03} stellar
population synthesis model] line flux and error measurements.
We combine a number of methods to analyze the full spectroscopic sample of
blue ellipticals (E$+$pE types) 
in terms of SF, black hole growth, and lack of emission.
We employ the standard \citet[][hereafter BPT]{baldwin81}
diagnostic for optical emission-line galaxies. In addition,
we identify spectroscopically
quiescent galaxies following \citet{peek10}, and
we use the ratio of [OII]$\lambda\lambda3726,3729$ to H$\alpha$
emission following \citet{yan06}
to identify the galaxies that are
neither quiescent nor have lines with sufficient S/N for the BPT diagnostic.
Finally, we discuss the low incidence of E$+$A galaxies identified by
\citet{goto07b}.

To compare the incidence of blue ellipticals with different emission properties
to control samples selected from the underlying galaxy population,
we analyze all DR4 galaxies meeting our stellar mass
($>10^{10}~h^{-2}~{\rm M}_{\sun}$) and redshift ($0.01<z\leq0.08$) cuts
that have good spectroscopic measurements; i.e., we remove 
low-S/N ($<5\,{\rm \AA}^{-1}$) spectra and a handful of objects with bad
flux measurements from pipe-line processing errors (e.g., discontinuities
in the spectral coverage) or stellar continuum fit errors. This results in a
spectroscopic sample of 58,455 galaxies with 92.1\% completeness, 
including 1267 (119) blue Es (pEs). The small
incompleteness is expected to be random given the SDSS spectroscopic
tiling \citep{blanton03a}.

\subsubsection{Blue Ellipticals in the BPT Diagram}
\label{sec:bpt}

\begin{figure}
\center{\includegraphics[scale=0.72, angle=0]{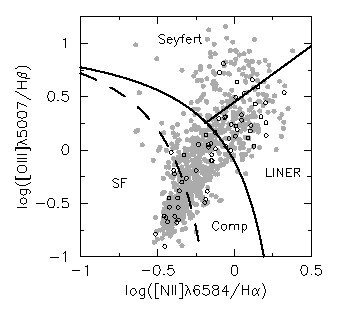}}
\caption[]{
The BPT \citep{baldwin81} diagnostic diagram for moderate-mass, blue
elliptical galaxies with normal (E, grey circles) and peculiar (pE, black
open circles) morphologies. All galaxies have well-detected emission 
line fluxes (with S/N$\geq3$) from MPA-JHU \citep{kauffmann03c,brinchmann04}.
The dashed curve shows the Kauffmann et al.
empirical division between pure starformers (SF) 
and galaxies
with SF-AGN composite (Comp) emission, although this interpretation is
controversial \citep{stasinska08}. The solid curve
shows the theoretical maximum starburst line derived by \citet{kewley01}.
The line sloping to the upper right is used by \citet{schawinski07b}
to separate Seyferts from LINERs.
\label{fig:BPT}}
\end{figure}

In Figure \ref{fig:BPT}, we show the BPT diagram
of [OIII]$\lambda5007$/H$\beta$ versus 
[NII]$\lambda6584$/H$\alpha$ line flux ratios for the
subset of our galaxies
with detectable nebular emission (line flux S/N$\geq3$) in
all four lines.
Following standard practices,
we use this diagnostic to classify galaxies into several emission types.
Galaxies with pure HII emission from SF (hereafter, SF type) lie below and
to the left-hand side of the dashed line defined empirically by 
\citet{kauffmann03c} to remove contamination from galaxies with
composite emission (Comp type) from both SF and an AGN.
\citet{kewley01} derived a theoretical maximum starburst line (solid curve)
to classify
galaxies dominated by AGN emission which lie above and to the right-hand
side of this line. Comp types lie between the dashed
and solid lines. 
Finally, we divide the Kewley et al. AGNs into 
Seyferts and LINERs following \citet{schawinski07b}. Recent work demonstrates
that LINER emission is typically spatially extended in present-day galaxies
\citep{annibali10,kehrig12,yan12},
which calls into question an AGN as the primary ionizing source
\citep[although some LINERs are clearly ionized by a central source,][]{pogge00},
rather ionization from hot post-AGB stars or white dwarfs is predicted
\citep[e.g.,][]{stasinska08}.
Therefore, the distinction between Seyferts and LINERs
is important when considering if any blue ellipticals
are being quenched by an AGN.
The interpretation of Comp types as SF-AGN composites is likewise
controversial. For example, these galaxies may be neither star-forming
nor an AGN, rather their
emission may be dominated by the same non-nuclear ionization sources
as many LINERs.

We compare the emission type demographics of visually-selected blue Es and pEs
against several control samples in Table~\ref{tab:2}.
With regard to the four BPT emission types (SF, Comp, Seyfert, LINER), we
find that the blue E and pE populations have comparable
relative type fractions, and we find that our sample has distinct
emission type percentiles from all three control samples:
red ETGs, blue LTGs and blue ETGs with visual disk features
(i.e., bulge-dominated spirals and S0 galaxies).
Our sample has a three to six times higher fraction of pure starformers than do
red ETGs, but a two to three times lower fraction compared to blue galaxies with disks.
Blue ellipticals have an incidence of Seyferts which is on par with that
of bulge-dominated blue disks and twice that of either red ETGs or blue LTGs.
It is interesting to note
that the presence of a substantial bulge in galaxies with cold gas (i.e., from
the blue cloud) is linked to a higher incidence of AGN activity as we expect
from the well-known correlation between black hole mass and bulge mass
\citep{ferrarese00,gebhardt00}.
In terms of LINER emission, blue ellipticals have a similar makeup as both red and blue
bulge-dominated populations. Finally, there is little difference between
the Comp fraction of blue ellipticals and the control samples, except for
the bulge-dominated blue disk galaxies which have nearly twice as many
Comps as their disk-dominated counterparts. 
Such a difference between late and early-type blue disk galaxies would be
expected if Comp emission included an AGN component. Perhaps, galaxies
with Comp emission are a mixed bag such that some are traditional SF-AGN
composites \citep{kauffmann03c}, while others may be `retired' 
\citep[i.e., non-star-forming][]{stasinska08} with non-circumnuclear
warm ionized gas \citep{yan12}. Given the ambiguous nature of Comp types,
we hereafter exclude these from AGN subsets. We consider only galaxies 
with Seyfert emission as definite AGNs, which
provides a lower limit of 5--10\% for the incidence of blue ellipticals
with an AGN, and we include LINERs for an upper limit to AGN fractions
with the understanding of their controversial nature. 
Overall, blue ellipticals are 50\% more
likely to have strong BPT emission (i.e., sufficient S/N in all four lines)
than red ETGs. Clearly, our selection
has identified the most active population of elliptical galaxies.
Finally, an important caveat is that we use only
optical emission lines to determine the presence of an AGN.
Studies have shown that optical selection alone may miss an important fraction
of AGNs detected at other wavelengths (e.g., radio or X-ray), and that
no single method will produce a sample that is both complete and reliable
\citep{barmby06}. We are undertaking a multiwavelength study of blue
ellipticals that is beyond the scope of this work.

\subsubsection{Spectroscopically Quiescent Blue Ellipticals}
\label{sec:quies}

To identify which of the remaining blue ellipticals with good spectroscopic
data have no detectable emission (i.e., quiescent), and distinguish these
from non-quiescent galaxies with emission too weak to be diagnosed with
the BPT diagram, we follow the method of
\citet{peek10}. Briefly, we calculate H$\alpha$ and [OII] equivalent 
widths (EWs) and their formal errors from the MPA-JHU line fluxes and
uncertainties. Peek \& Graves found an ellipse, centered on
([OII] EW, H$\alpha$ EW)$=(0.23,0.18)$ with semiaxes of $4.0\,{\rm \AA}$ in
[OII] and $0.94\,{\rm \AA}$ in H$\alpha$, encloses most quiescent galaxies
with a minimum contamination from low-level star-formers and LINERs.
Moreover, they required that 
the formal errors of the [OII] and H$\alpha$ line strengths of each
quiescent galaxy be included within the same ellipse.
Additionally, we include galaxies in our quiescent selection that  
have no detectable H$\alpha$ line
(flux S/N$<3$), or detectable but weak (H$\alpha$ EW $\leq2 {\rm \AA}$)
H$\alpha$ emission. These cuts ensure that we analyze the small subset
of low-redshift galaxies with good spectra but lacking the 
[OII]$\lambda\lambda3726,3729$ doublet owing to the effective 
$3800-9200\,{\rm \AA}$ coverage of the SDSS spectrographs.
We tabulate the incidence of quiescent galaxies among blue ellipticals
and several control samples in Table~\ref{tab:2}.

We find similar incidences ($\sim40\%$) of blue E and pE galaxies are
spectroscopically quiescent.
This fraction is significantly less than for red ETGs and an order of
magnitude higher than the quiescent portion of blue LTGs.
The inactive fraction of blue ellipticals is also much higher than
that for blue ETGs with disk features.  
These galaxies have sufficiently high-S/N spectra
($98.6\%$ have median $S/N>10\,{\rm \AA}^{-1}$) to verify that
the quiescent fraction clearly lacks emission from SF or an AGN.
The lack of activity coupled with unusually blue colors is suggestive of a
recently quenched population of elliptical galaxies.
As outlined in the Introduction, such galaxies are consistent with the
predictions of gas-rich major merging and may represent strong candidates
for first-generation ellipticals.  We focus on the nature of this interesting
subpopulation in \S~\ref{sec:Quenched}.

\subsubsection{Blue Ellipticals with Weak H$\alpha$ and/or [OII] Emission}
\label{sec:weakEm}

About 5\% of galaxies with good spectroscopic data meet neither
the BPT diagnosis criteria nor the quiescent selection; i.e., they have
either weak emission lines or strong emission in only a few lines like
H$\alpha$ and [OII].
We split these objects into two `weak-emission' types following \citet{yan06}.
First, `LINER-like' (hereafter, weak-LINER)
systems have H$\alpha$ line flux S/N$\geq 3$, [OII] EW S/N$\geq 3$, and
high [OII]/H$\alpha$ EW ratios such that
[OII] EW$>5({\rm H}\alpha$ EW)$-7$. Secondly, weak-SF types
have low [OII]/H$\alpha$ ratios ([OII] EW$\leq 5({\rm H}\alpha$ EW)$-7$),
or no detectable [OII] (EW S/N$<3$) and H$\alpha$ EW$>2\,{\rm \AA}$.
The latter H$\alpha$ EW criterion 
empirically matches the H$\alpha$ EWs of the low-[OII]/H$\alpha$ sample.
The number of weak-SF galaxies in our sample is statistically zero
and we exclude them from further analysis.
In what follows, we find that weak-LINER and `normal' (BPT) LINER galaxies
behave the same in parameter spaces that we explore. We test and find that
including/excluding weak-LINERs does not change our results. For completeness, 
we elect to combine all LINERs into one subset, but we show their separate
contributions in Table~\ref{tab:2}.

\begin{table*}
\caption{Demographics of galaxy emission types.}
\label{tab:2}
\begin{tabular}{lccccccc}
\hline\hline
Sample & $N_{\rm spec}$ & SF$^{\rm a}$ & Comp$^{\rm a}$ & Seyfert$^{\rm a}$ & LINER$^{\rm a}$ & weak-LIN$^{\rm b}$ & Quies$^{\rm c}$\\
\hline
Blue E      &  1267  &  17$\pm1$ &  22$\pm1$ &  8$\pm1$ &  12$\pm1$ &  5$\pm1$ & 36$\pm2$ \\
Blue pE     &  119   &  13$\pm3$ &  15$\pm4$ &  5$\pm2$ &  19$\pm4$ &  3$\pm2$ & 43$\pm6$ \\
\hline
Red ETG    & 29554   &   3 &  14 &  3 &  16 &  4 & 57 \\
Blue LTG   & 15343   &  51 &  21 &  3 &   7 &  1 &  3 \\
Blue ETG disk & 2684 &  31 &  38 &  7 &  14 &  2 &  6 \\
\hline
\end{tabular}
\vskip 8pt
\begin{minipage}{\hdsize}
Percentiles of emission types for our selection of blue
ellipticals
and three control samples from galaxies with good spectroscopic data.
We provide $\sqrt{N}$ errorbars for the blue elliptical subpopulations;
for reference, the counts among all emission types for the control samples
are such that $\sqrt{N}$ errorbars are $<0.5\%$.
We note that summing the emission percentiles for each sample will not add 
precisely to 100\% owing to the omission of weak-SF types 
(see \S~\ref{sec:weakEm}).
For example, 13\% of blue LTGs are this type, but for all other
types the percentage is $0-3\%$. \\
$^{\rm a}$ Based on the BPT diagnostic diagram (Fig.~\ref{fig:BPT}) and 
classifications following \citet{kewley01}, \citet{kauffmann03c} and
\citet{schawinski07b}; \S~\ref{sec:bpt}. \\
$^{\rm b}$ Based on the weak-emission diagnostic of \citet{yan06};
\S~\ref{sec:weakEm}.\\
$^{\rm c}$ Based on the spectroscopically-quiescent definition of \citet{peek10};
\S~\ref{sec:quies}.
\end{minipage}
\end{table*}

\subsubsection{How Many Blue Ellipticals are E$+$A Galaxies?}

A number of studies have linked post-starburst galaxies with major gas-rich
mergers and the formation of new ellipticals 
\citep[e.g.,][and references therein]{yang06,goto07a,snyder11}.
Therefore, any blue ellipticals that are the recent end product of a
starburst-producing merger should exhibit post-starburst or so-called
E$+$A signatures.  E$+$A (or K$+$A) galaxies have spectra that are a
combination of an ETG (K-type) stellar continuum and strong Balmer absorption
from a significant population of A-type stars that were formed in a brief
burst.
We identify 35 E$+$A galaxies from SDSS DR5 \citep{goto07b}
within our stellar mass-limited sample of 58,455 galaxies
with $0.01<z\leq0.08$ and spectroscopic data. 
This subset represents only 6\% of the full Goto catalog,
the bulk of which have higher redshifts and/or lower masses. 
We find that the tiny fraction of high-mass, low-redshift E$+$A galaxies mostly
have blue-cloud colors and high concentrations; i.e., 27 meet our blue ETG
selection. These break down into
22 with high classifier agreement (8 E,
3 pE, 1 SPM, 8 U, and 2 iD) plus five with lower classification agreement, yet
combinations of plausible post-merger and U classifications.
The low fraction with clear morphological peculiarities (4 of 22) emphasizes
that the dynamical and stellar population `clocks' of merger remnants
are different \citep{gyory10}. Given the typical time-scale for post-merger
disturbances \citep[$<0.5$\,Gyr,][]{lotz08b,lotz10a,lotz10b},
the low disturbed fraction calls into question predictions of
similarly short E$+$A phase lifetimes \citep{snyder11}.
In summary, very few high-mass blue ellipticals at low redshift ($<0.1\%$)
have experienced a recent starburst\footnote{We find similarly small E$+$A
fractions among high-mass blue SPMs and blue ETGs
with uncertain morphologies.}. If starbursts
are ubiquitous with gas-rich major merging, then this result suggests that only a tiny
fraction of blue ETGs with plausible post-merger morphologies 
(from normal E to SPM) are actually
post-mergers. Yet, as discussed in the Introduction, the frequency of
gas-rich interactions with 
${\rm M}_{{\rm gal,}{\star}}\geq10^{10}~h^{-2}~{\rm M}_{\sun}$ is 
much higher than that of similarly massive post-starburst galaxies.
Therefore, it is probable that a larger fraction of blue 
ellipticals are recent major-merger remnants,
but only a subset of such mergers produce strong starbursts
\citep{robaina09}. In \S~\ref{sec:QEhow},
we discuss the E$+$A incidence among RQEs
and the implications for the starburst phase in gas-rich merging at
late cosmic times.

\subsection{Color-Sigma}
\label{sec:colors}

\begin{figure*}
\center{
  \includegraphics[scale=0.7, angle=0]{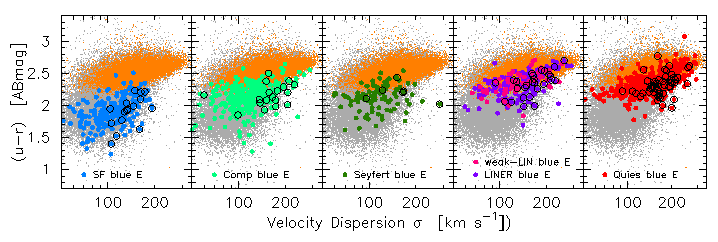}
}
\caption[]{
Color-sigma distribution for spectroscopic sample of nearby ($0.01<z\leq0.08$),
massive (${\rm M}_{{\rm gal,}{\star}}\geq10^{10}~h^{-2}~{\rm M}_{\sun}$)
galaxies. Colors are from SDSS model magnitudes (modelMag) 
corrected for Galactic extinction and
$K+$evolution corrected to $z=0$. Velocity dispersions (velDisp) are  
measured from the SDSS fiber spectrum. Each panel shows all 
galaxies in grey, and a control sample of 16,976 spectroscopically-quiescent,
ETGs ($c_r\geq2.6$) with $(g-r)$ red-sequence colors in
yellow. From left to right, the panels contain different subsets
of blue elliptical galaxies based on emission types: pure star-forming 
(SF, blue), composite emission
(Comp, cyan), Seyfert (green), LINER from the BPT diagram (purple),
weak-emission LINER (pink), and quiescent (red). Open black circles designate
blue ellipticals with peculiar morphology.
\label{fig:colorSigma}}
\end{figure*}

We compare the color versus velocity dispersion distributions of blue elliptical
galaxies split into five emission types (SF, Comp, Seyfert, LINER and 
quiescent) in Figure~\ref{fig:colorSigma}. For each galaxy with good
spectroscopic data (see \S~\ref{sec:EmTypes}), we plot the
$(u-r)$ color corrected to rest-frame $z=0$ as a function of 
$\sigma$ (a proxy for dynamical mass) measured from the SDSS fiber spectrum.
These choices provide a greater dynamic
range for exploring the color and mass distribution of our sample.
In each panel, we provide the underlying sample of all spectroscopic galaxies
and we highlight a control sample of
16,976 spectroscopically-quiescent, red ETGs which represent the red-sequence
in the $^{0.1}(g-r)$ color versus stellar mass space that we use to
define our blue elliptical color selection (Fig.~\ref{fig:blueseln}).

A number of interesting trends are immediately clear in 
Figure~\ref{fig:colorSigma}. Foremost, we find that our sample of
visually-selected
ellipticals with unusually blue optical colors follows a similar
sequence in color, mass and emission activity as 
\citet[][hereafter, S07]{schawinski07b}
found for a broad selection of ETGs; i.e., visually classified E/S0 
galaxies with {\it no} cuts in mass nor color.
In accord with their results, we find that
star-forming blue ellipticals populate the lower-mass
blue cloud, while those with composite and Seyfert emission
tend to focus in the so-called green valley, and LINER and quiescent systems
favor higher masses and redder colors. A detailed comparison with S07
(their Fig.\,7)
reveals a number of minor differences that are
explained by differences in sample selection.
First, our stellar mass cut results in fewer SF types at 
$\sigma<100$\,km\,s$^{-1}$ compared to the SF types in S07 which
make up more than half of the low-$\sigma$ ETGs. Even still,
the subpopulation of blue ellipticals with active SF have the lowest
average velocity dispersion among all other emission types in our
stellar mass-limited sample.
More importantly, our blue $(g-r)$ color cut preferentially excludes the
redder half of the red sequence. This is seen clearly in the two right-most
panels of Figure~\ref{fig:colorSigma}, in which the LINER and quiescent
subsamples of blue ellipticals 
populate only the bluer portion of the $(u-r)$ quiescent
red sequence (in orange). It is important to note that our results are 
fully consistent with those of S07, who found that LINER and quiescent ETGs
populate nearly identical regions of color-sigma space, which is a close
match to our red sequence defined by the control sample of 
spectroscopically-quiescent red ETGs.
By selecting ellipticals with unusually blue
colors, we are merely focusing on the bluest examples of LINER 
and quiescent types.
This selection effect also holds true for the Comp and Seyfert blue ellipticals
which S07 showed have 
green-valley average colors, but also extend well into the
red sequence.

A striking feature of Figure~\ref{fig:colorSigma} is the concentrated
location of blue pEs in each panel.
Visually-selected blue ellipticals
with evidence of recent tidal activity appear to
prefer the high-velocity dispersion
and bluer edge of the color-sigma distribution for all emission types
except Seyferts. For Comp, LINER and quiescent emission types, the blue pE
concentration aligns with the green valley and provides evidence in favor
of these galaxies being in active transition between the blue cloud and
the red sequence. Likewise, both blue elliptical subpopulations (E and pE)
with Seyfert emission prefer this transition region.  
It is tantalizing to speculate that gas-rich merging,
whether minor or major, is playing an important role in defining the
massive edge of the blue cloud. A popular hypothesis for defining the mass
limit of blue-cloud galaxies is that star-forming
galaxies {\it may} grow in mass until some mechanism
quenches them and sends them to the red sequence 
\citep[e.g.,][]{bell04b,faber07}. This critical mass appears to correspond
to $\sigma\sim 200$\,km\,s$^{-1}$. The color-sigma distribution of quiescent
blue pEs is particularly compelling in light of such evolutionary scenarios.
These galaxies concentrate within a tight locus with green-valley
colors and velocity dispersion just below and up to the 
critical mass limit. Indeed, nearly all quiescent blue ellipticals with 
$\sigma >200$\,km\,s$^{-1}$ have normal morphologies and red-sequence
$(u-r)$ colors, while most with $\sigma <140$\,km\,s$^{-1}$ likewise lack
morphological peculiarities but tend toward bluer colors.
We note that this result is tentative given the uncertainty in our ability
to detect asymmetric features in low-mass galaxies
($\leq 3\times 10^{10}~h^{-2}~{\rm M}_{\sun}$) at all redshifts
that we study (see \S~\ref{sec:classTests}).

Further dissecting our sample of blue ellipticals in terms of spectroscopic emission
reveals a number subpopulations experiencing either different mass assembly
processes or different phases of a similar process.
For example, the low-$\sigma$ blue ellipticals with active SF may be
examples of rejuvenated ETGs identified by \citet{thomas10}.
Or these galaxies may be the recent end product of a major merger-triggered
SF event such as a pre-existing elliptical accreting a large gaseous
companion or a spiral-spiral merger. The latter case would represent
an important population of pre-quenched first-generation 
ellipticals, and their 
moderately-relaxed (pE) to fully-relaxed (E) appearance would imply
an extended period
of merger-induced SF that lasts significantly longer than the strong
morphological asymmetries identifiable for $<0.5$\,Gyr in new remnants
\citep{lotz08b,lotz10a,lotz10b}. We note that these star-forming ellipticals are
likely a subset of the blue ETGs (E/S0's) studied in \citet{schawinski09c}.
Additionally, the blue ellipticals with
Comp emission have colors and emission that are commonly identified
with a transitory nature between star-forming and AGN,
and also a quenching population that is migrating redwards
\citep[e.g.,][]{schawinski07b}. Yet, the properties of these ellipticals
are also consistent with red-to-blue (or red-to-green valley) evolution from
a minor frosting of new stars. Needless to say, these two
subpopulations 
require closer scrutiny and will be the
subject of forthcoming papers. 
For the remainder of this paper, we will focus on isolating the best examples
of recently quenched (non-star-forming) ellipticals and exploring their
nature.

\section{Recently Quenched Ellipticals}
\label{sec:Quenched}

Among the emission types of blue ellipticals identified
in the previous section, the Seyfert and quiescent (non-emission)
subpopulations are particularly noteworthy in the context
of the gas-rich merging scenario.
The green-valley colors of the Seyferts suggest they have either
low-level, on-going SF or recently quenched SF. In what follows, we use 
the optical color-color plane to further constrain the non-star-forming
population and identify the subset that have likely been quenched by
the AGN. Moreover, the quiescent
blue ellipticals may have experienced a recent episode of SF that has been
subsequently quenched, or it is possible that these galaxies are merely
old red ellipticals scattered from the red sequence into our blue $(g-r)$
color selection. As such, we use
stellar ages to identify a new population of very young but non-star-forming
-- recently quenched -- elliptical galaxies and distinguish these from
systems with old stars.
We then define our sample of RQEs,
characterize their nature, and discuss
possible quenching mechanisms.

\subsection{Identification}

To confirm the non-star-forming nature of blue ellipticals with no emission
and to identify which Seyfert systems lack SF, we
analyze their location in $(u-r)$ versus $(r-z)$ space.
\citet[][hereafter H12]{holden12} demonstrated
that SDSS photometry robustly separates
{\it passive-red}, non-star-forming galaxies
from {\it dust-reddened} star-formers
in the spirit of similar work using optical plus near-IR data
\citep[$UVJ$; e.g.,][]{labbe05,wuyts07,williams09,brammer11,patel12}. 
Using a sample of SDSS galaxies with very similar selection to ours,
H12 found an optimum boundary defined by 
$(u-r)>2.26$, $(r-z)<0.75$, and
$(u-r)>0.76 + 2.5(r-z)$, that contains the highest fraction (0.82)
of galaxies without detectable H$\alpha$ emission (their quiescent, early-type
galaxies) and
with the lowest contamination (18\%) of H$\alpha$ emitters (their star-forming,
late-type galaxies).
We follow the prescription of H12
and use SDSS model magnitudes (modelMag)
corrected for Galactic extinction and
$K+$evolution corrected to $z=0$. 
In Figure~\ref{fig:urz_control}, we show how well the H12 boundary 
distinguishes two control samples selected from our emission-type definitions:
(i) quiescent red ETGs, and (ii) blue LTGs with pure SF emission.
Compared to what H12 reported based on
their simple H$\alpha$-based definitions, we find even better separations
such that 91.6\% (94.1\%) of all quiescent systems (red ETG subset)
are contained within the H12 non-star-forming
region, while 97.0\% (99.0\%) of all pure
starformers
(blue LTG subset) fall outside this boundary.
We find similar results using colors based on the SDSS fiber magnitudes;
we opt to use total integrated colors as H12 did so that we can be
fairly confident that non-star-forming galaxies lack SF in their outskirts.

Additionally in Figure~\ref{fig:urz_control}, 
we explore the $(u-r)$--$(r-z)$ dependencies on light-weighted
stellar ages $\log_{10}(t_{\rm age}/{\rm yr})$ and stellar metallicities
$\log_{10}(Z/Z_{\odot})$ from \citet{gallazzi05}.
The ages and metallicities represent the median-likelihood estimates
computed from stellar population model fits to the SDSS
fiber spectroscopy for galaxies with high-quality (median 
S/N/pixel$>20$) spectra. The Gallazzi et al. age and metallicity
estimates are constrained to within $\pm 0.15$\,dex for most galaxies.
Galaxies cannot be parametrized by a single age, rather light-weighted ages
are a composite of multiple ages and are very sensitive to SF within the
last Gyr \citep{serra07}, which makes these ages useful indicators of
recent SF and the presence of younger stars. 
For the full spectroscopic sample we analyze, 92.8\%
have good age and metallicity estimates -- this completeness holds for
blue ellipticals and all comparison samples.
We find that galaxy colors in the non-star-forming region correlate with both 
stellar age and
metallicity in the sense that redder (in both
$u-r$ and $r-z$ colors) quiescent red ETGs are older and more metal-rich.
Star-forming blue LTGs also follow a scattered
age--$(u-r$) color relationship such that the bluest galaxies have the
youngest ages, but these galaxies span the full range in metallicities
with no clear trends with color.
Color trends with age and metallicity for ETGs \citep{gallazzi06}
and with age for all galaxies \citep{chilingarian12}
are well documented.

With the $urz$-age-metallicity trends for quiescent\footnote{Hereafter,
quiescent refers to spectroscopically devoid of emission lines 
(see \S~\ref{sec:quies}), while non-SF refers to non-star-forming colors
in $urz$ space.} and pure star-forming control
samples in mind,
we plot the color-color space of the quiescent and pure star-forming 
subpopulations of blue ellipticals in the left panel of
Figure~\ref{fig:urz_RQE_seln}.
In striking contrast to quiescent red ETGs which are very well bounded
by the H12 non-SF region, one-quarter of the quiescent blue elliptical
subpopulation extends bluewards of the $(u-r)=2.26$ boundary.
These quiescent objects are remarkably young with an average
stellar age of 3.3\,Gyr. 
While the H12 non-SF region does not capture the
youngest portion of quiescent blue ellipticals, it does robustly 
distinguish the star-forming subset; we find 97.4\% of blue ellipticals
with SF emission outside of the H12 region. Moreover, bluewards of
the H12 $(u-r)=2.26$ line, the quiescent and SF subsamples remain distinct.
Therefore, we define a {\it modified non-SF region} that extends the H12 
diagonal boundary down to a bluer $(u-r)=1.90$ cut
that fully encompasses our sample of quiescent blue ellipticals.
We note that the extended portion of our modified non-SF region has a very
low (1.8\%) contamination from star-forming blue ellipticals.
The galaxies within our modified non-SF region span a large range
in $t_{\rm age}$ from 1--10\,Gyr. For comparison, we show the mean and
standard deviation of the colors for quiescent red ETGs with
$t_{\rm age}<5$\,Gyr and $t_{\rm age}>9$\,Gyr.
We highlight the disproportionate number of quiescent pEs with young,
non-SF colors which suggests a link between tidal activity and recent
quenching of SF.

Given the unique $urz$ colors of the young quiescent subset of blue ellipticals,
we turn our attention to the Seyfert subpopulation to see if there are any
equally young, non-star-forming AGNs.
Under the assumption that the nebular emission of an AGN is distinct from
the stellar continuum that produces the global colors in the host galaxy,
we investigate the $urz$ colors of the Seyfert subpopulation of our
sample in the middle panel of
Figure~\ref{fig:urz_RQE_seln}.
A majority of these galaxies fall within our modified non-SF boundary, and most
have young stellar ages (average $t_{\rm age}=4.0$\,Gyr) and correspondingly 
bluer $(u-r)$ colors in the same manner as the young quiescent examples. 
This result shows that some blue ellipticals with
Seyfert emission have had SF recently quenched, possibly by the active AGN.
We likewise investigate the $urz$ colors of blue ellipticals with LINER
emission in the right panel of Figure~\ref{fig:urz_RQE_seln}).
We combine the subpopulations of strong (BPT) and weak \citep{yan06} LINERs
and find that nearly 90\% lie within our modified non-SF region. 
Moreover, nearly 1/4 of the non-star-forming LINERs 
populate the very-blue $(u-r)<2.26$ space 
and have even {\it younger} average ages (2.7\,Gyr) than the quiescent
subpopulation below this color cut. It is possible
that some LINERs, whether their ionization source is actually a low-powered 
AGN or an extended source of hot post-AGB stars, play a role in quenching
SF in elliptical galaxies. Therefore, we include young, non-SF LINERs
in our analysis of RQEs
and we explore the frequency of possible AGN quenching
in more detail in \S~\ref{sec:Qagn}. 
Finally, we check the $urz$ colors and ages of Comp emission types and
find that a majority (61\%) have {\it star-forming} colors. The subset
that lie within our modified non-SF region tend to have colors that straddle
the diagonal non-SF/SF boundary. As such, these objects appear to be
transitory in terms of their $urz$ colors as well as their optical emission
lines. While 20\% of the Comps have very-blue $(u-r)<2.26$ colors, their
average age is 4.0\,Gyr, which is much older than the other emission-line
systems with similar colors. For these reasons, and the controversial nature
of whether Comps are star-forming or not (see \S\,\ref{sec:bpt}), 
we opt to exclude them from a
recently quenched selection. We confirm that their inclusion would not change
our results significantly.

\subsubsection{RQE Definition}
\label{sec:QEidentif}
The elliptical galaxies with unusually blue but non-star-forming colors,
very young light-weighted stellar ages, and no detectable emission
are clearly examples of ETGs that were forming stars, experienced a recent
quenching of SF, and are now transitioning to the red sequence.
Additionally, similarly young and non-star-forming blue 
ellipticals with Seyfert or LINER
emission have all the hallmarks of recent SF quenching, possibly as a result
of the AGN or LINER emission.
Therefore, we select RQEs among blue ellipticals using the following
simple criteria: (i) $urz$ colors within
our modified non-SF region, (ii) stellar ages
of $t_{\rm age}\leq3$\,Gyr, and (iii) either spectroscopically quiescent
or with Seyfert or LINER emission.
The RQEs so defined are shown in green in Figure~\ref{fig:urz_RQE_seln}.
We choose this age cut because $<1\%$ of
quiescent red ETGs are this young, compared to 24\% of quiescent blue
ellipticals. We note that the youngest red ETGs are 0.14\,mag redder
in $(u-r)$ and 0.06\,mag redder in $(r-z)$ than the RQEs. This
suggests that equally young red ETGs and RQEs 
have different stellar populations.
For example, very young red ETGs, in a light-weighted sense,
may be examples of ETGs with
low-level recent SF identified by \citet{kaviraj07c} and others using GALEX
plus optical data. These galaxies experienced a sprinkling of 1--3\% new
stars by mass very recently ($<1\,{\rm Gyr}$), while the RQEs we 
identify may have a larger fraction of blue stars a bit older than 1\,Gyr.
The low incidence of E$+$A galaxies (see Table~\ref{tab:4}) 
limits the mass fraction of new stars added in RQEs. The bottom line is that
there are virtually zero red-sequence ETGs with the ages {\it and} colors
of the RQEs; the RQEs are a unique subpopulation of ETGs.
Moreover, while similarly young and blue subpopulations of blue ETGs with
disks and blue LTGs exist, they comprise a small fraction (2.5\% and 1.1\%,
respectfully) of their kind. 
We stress that RQEs are not the result of aperture bias; i.e., they have the
same redshift distribution as the full stellar mass-limited sample from
SDSS DR4. A total of 172 blue E and pE visual types meet our RQE
definition. We present a sample of the RQE catalog in Table~\ref{tab:samp};
the full catalog is available electronically.

\subsubsection{Testing A Pure Color-Color Selection of Recently Quenched ETGs}
A number of recent studies have used the rest-frame $(U-V)$ versus $(V-J)$
color-color diagram to identify high-redshift ($1<z<3$) quiescent
(i.e., non-star-forming) galaxies
to study their cosmic evolution \citep{williams09,wuyts09b,bell12}.
\citet{whitaker12a} have taken this approach one step further by separating
massive quiescent galaxies at 
$0.2<z<2.0$ into young (recently quenched) and old to compare their size
and number density evolution.
These authors selected
$>5\times 10^{10}\,{\rm M}_{\sun}$ galaxies from the NEWFIRM Medium-Band Survey,
identified the 15\% bluest quiescent systems using a rest-frame
$(V-J)$ color cut in the $UVJ$ plane, and found that their average age based
on their composite rest-frame spectral energy distribution is roughly
one-half that of the older quiescent galaxies. This work demonstrates
the utility of the correlation between unusually blue non-SF colors and
unusually young stellar ages discussed in the previous section
for selecting recently
quenched ETGs from large galaxy samples lacking spectroscopic data.
A key question remains as to what mechanism(s) is(are) shutting off
SF in these systems. Applying a similar technique to
isolate the youngest quenched systems
could prove powerful for constraining quenching processes in ETGs 
with new samples that have sufficient numbers of massive
galaxies with good rest-frame $UVJ$ data.

\begin{figure*}
  \center{
  \includegraphics[scale=0.47, angle=0]{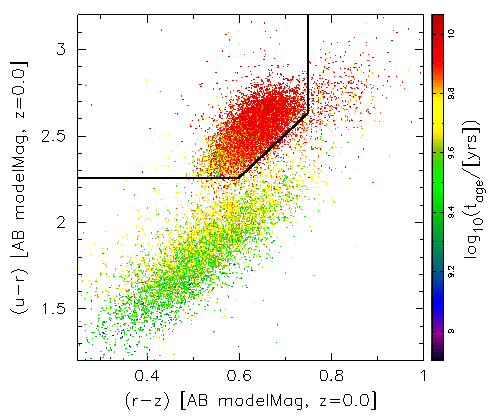}
  \hspace{0.9cm}
  \includegraphics[scale=0.47, angle=0]{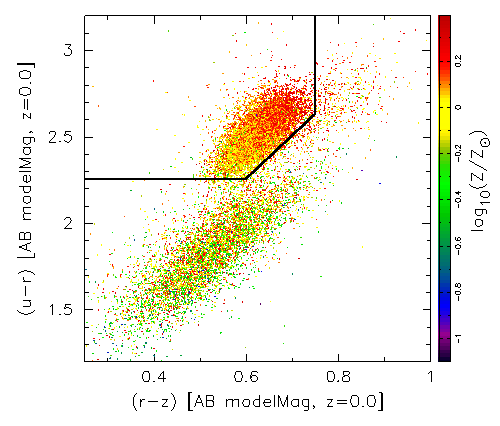}
  }
\caption[]{
Bimodal color-color distribution of spectroscopically-quiescent, red-sequence
ETGs (inside boundary) and pure star-forming (BPT HII emission), 
blue-cloud LTGs (outside
boundary). The H12 boundary (black lines)
efficiently selects galaxies
without detectable H$\alpha$ emission (their quiescent, ETG definition).
The $(u-r)$ and $(r-z)$ colors are based on SDSS
model magnitudes (modelMag) corrected for Galactic extinction and
$K+$evolution corrected to $z=0$. 
The galaxy data points in each panel are color-coded by the light-weighted
median stellar age and median
metallicity ({\it right}) estimates
derived by \citet{gallazzi05}.
\label{fig:urz_control}}
\end{figure*}

\begin{figure*}
  \center{\includegraphics[scale=0.63, angle=0]{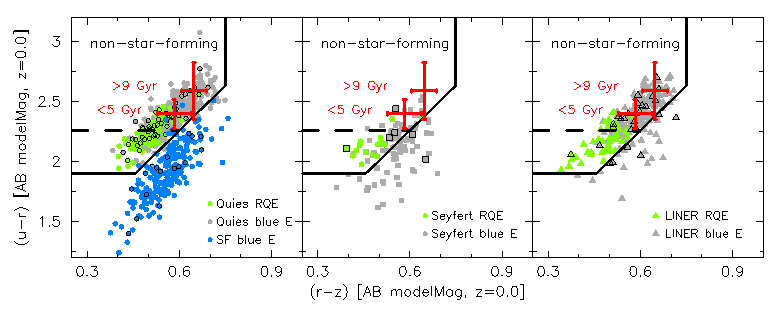}}
\caption[]{Color-color distributions for RQEs
(green data) and non-RQE blue ellipticals (grey)
subdivided by spectroscopic type:
({\it left}) quiescent RQEs and older quiescent blue ellipticals with
pure star-forming blue ellipticals for comparison shown in blue; 
({\it middle}) Seyferts; and
({\it right}) LINERs, both strong (BPT) and weak \citep{yan06} emission types.
The data in each panel are as in
Fig.~\ref{fig:urz_control}. The horizontal boundary of the H12
non-star-forming region is shown with a dashed black line.
Our `modified' non-star-forming region (solid black lines) extends to
$(u-r)=1.9$. Peculiar (pE) ellipticals for each subpopulation
are shown with open symbols.
The average colors of spectroscopically-quiescent, red ETGs with stellar
ages $\leq5$\,Gyr and $>9$\,Gyr are plotted in red with 
one standard deviation errorbars.
\label{fig:urz_RQE_seln}}
\end{figure*}

\begin{figure}
  \center{\includegraphics[scale=0.75, angle=0]{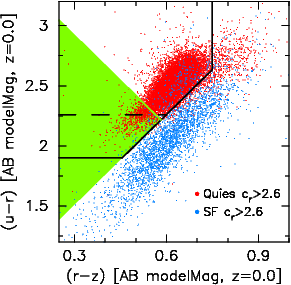}}
\caption[]{Color-color selection of recently quenched ETGs (green shaded
triangle). ETGs are automatically selected using an $r$-band
central light concentration cut of $c_r\geq 2.6$. We plot all ETGs that
are spectroscopically quiescent (red) and pure starformers (blue).
The modified and H12 non-star-forming regions are as in
Fig.\,\ref{fig:urz_RQE_seln}. The recently quenched ETG selection region
is empirically defined by extending the non-SF diagonal line and
encompassing 5\% of all non-SF ETGs.
\label{fig:urz_empirical}}
\end{figure}

Here, we provide a valuable consistency check to the work of 
\citet{whitaker12a} by testing a similar pure, color-color selection of 
the youngest, non-star-forming ETGs
using 35,115 high-mass ($>\geq10^{10}~h^{-2}~{\rm M}_{\sun}$) SDSS galaxies
with automated ETG morphologies ($c_r\geq 2.6$), spectroscopic emission types
(\S\,\ref{sec:EmTypes}), and good light-weighted stellar age estimates
from \citet{gallazzi05}. These data allows us to quantify the ages and
emission activity of galaxies meeting the simple
selection criteria illustrated in Figure~\ref{fig:urz_empirical}.
We exploit the fact that the $urz$ (H12) and $UVJ$ \citep{williams09} 
identifications of
non-star-forming galaxies each consist of a polygon with a diagonal boundary
that both divides and runs parallel to the two (SF and non-SF) sequences.
Therefore, we extend the diagonal boundary of the non-SF region bluewards
indefinitely such that all galaxies above this line and to the left of
$(r-z)=0.75$ are considered non-star-forming. This region contains 25,900
high-mass
ETGs with very low (1\%) contamination from galaxies with pure SF emission.
The analogous criterion using
rest-frame $UVJ$ colors would be above the diagonal boundary and to the left
of $(V-J)=1.6$.
Next, we use a line perpendicular to the SF/non-SF diagonal to isolate
the 5\% bluest non-SF ETGs; here, the line is 
$(u-r)=3.67 - 2.5(r-z)$. The criteria in $UVJ$ space would be a likewise
perpendicular line to the corresponding SF/non-SF diagonal positioned
to select a given percentages of the bluest non-SF galaxies. 
Since the non-SF colors correlate with light-weighted
stellar age (Fig.~\ref{fig:urz_control}), the perpendicular line is roughly
a straight age cut. This cut could be modified to select a larger or
smaller percentage of the non-SF population. The important thing is that
this empirical selection identifies the youngest non-SF population; i.e.,
the green triangular region shown in
Figure~\ref{fig:urz_empirical} selects the most recently quenched ETGs.

The average light-weighted stellar 
age of galaxies meeting this pure color-color selection is
4\,Gyr, with one-third having $t_{\rm age}\leq3$\,Gyr and only
one-quarter with ages older than 5\,Gyr. Put another way, the bluest 5\%
of $>10^{10}\,{\rm M}_{\sun}$ non-SF ETGs at low redshift have an
average age that is about one-half that of the older non-SF population.
This selection
identifies 62\% of all non-SF galaxies younger than 3\,Gyr
and 82\% of the RQEs we selected in the previous section.
We note that more than half of the galaxies selected in this manner
have no detectable emission lines, while another one-third
are split evenly between galaxies with LINER and composite emission.
Seyferts make up 7.6\%, with 40\% of these having $t_{\rm age}\leq3$\,Gyr.
In short, using a pure optical-near-infrared color selection of
recently quenched ETGs based on the SDSS, for which we can fully 
exploit spectroscopic 
estimates of age, we confirm the technique of \citet{whitaker12a}.

\begin{table*}
\caption{Example from the catalog of RQEs.}
\label{tab:samp}
\begin{tabular}{lcccccc}
\hline\hline
NYU ID & R.A. & Dec & $z$ & ${\rm M}_{{\rm gal,}{\star}}$ & Type & Emission \\
 (1) & (2) & (3) & (4) & (5) & (6) & (7)\\
\hline
nyu246089 & 7.15144084 & $-$9.93114501 & 0.050301 & 10.088 & E & LINER \\
nyu27832 & 7.60293981 & $+$0.15565728 & 0.060158 & 10.048 & E & LINER \\
nyu271516 & 9.76905203 & $-$9.31515332 & 0.065592 & 10.019 & E & Quies \\
nyu106052 & 11.82478667 & $+$15.54226519 & 0.07973 & 10.236 & E & Seyfert \\
nyu107049 & 12.01295453 & $+$16.01610727 & 0.050202 & 10.604 & E & Quies \\
nyu99345 & 13.31444145 & $+$14.8610745 & 0.06056 & 10.107 & E & LINER$^{\rm a}$ \\
\hline
\end{tabular}
\vskip 8pt
\begin{minipage}{\hdsize}
Col. (1): galaxy
identification number \citep[from the DR4 NYU-VAGC,][]{blanton05}.
Cols. (2)--(3): Epoch J2000.0 celestial coordinates in degrees
from the SDSS. Col. (4): 
SDSS spectroscopic redshift from the NYU-VAGC.
Col. (5): Stellar mass estimates in units of
$\log_{10}(h^{-2}{\rm M}_{\sun})$ based on SDSS Petrosian photometry
and \citet{bell03b} M/L ratios.
Col. (6): Visual classification type (see \S~\ref{sec:VisScheme} for details).
Col. (7): Optical emission type (see \S~\ref{sec:EmTypes} for details).
Table~\ref{tab:samp} is published in its entirety in the electronic edition
of the journal. A portion is shown here for guidance regarding its form
and content.\\
$^{\rm a}$ LINER based on the \citet{yan06} criteria.
\end{minipage}
\end{table*}

\subsection{Nature of RQEs}
\label{sec:QEnature}

Here we characterize this unique population of low-redshift, high-mass RQEs.
These galaxies have either no emission lines or are Seyfert/LINER types and
meet the simple $urz$ color and age criteria given in \S~\ref{sec:QEidentif}.
These objects make up only
0.32\% of the full DR4 sample with $0.01<z\leq 0.08$ and
stellar masses between $10^{10}$ and $10^{11}~h^{-2}~{\rm M}_{\sun}$.
The number density of RQEs is $2.7\times 10^{-5}\,h^3\,{\rm Mpc}^{-3}$, which
represents a lower limit owing to the fact that we have limited our
analysis to blue ellipticals with high classification agreement. 
We find an upper limit of
$4.7\times 10^{-5}\,h^3\,{\rm Mpc}^{-3}$ for blue ETGs
without clear disk features by including galaxies with uncertain 
classifications that may be elliptical in morphology; i.e.,
high-agreement U types and E/U types (with two E and two U classifications).
A detailed
study of the nature of all SDSS galaxies meeting our recently quenched
definition is beyond the scope of this work, but we note that 
\citet{mendel13} identified a larger sample of recently quenched galaxies in 
the SDSS and found a distinct lack of disk-dominated systems.
These authors quote a higher number density of 
$\sim6.4\times 10^{-4}\,h^3\,{\rm Mpc}^{-3}$, which mainly reflects their lower
mass selection criteria of
$\log_{10}({\rm M}_{{\rm gal,}{\star}}/{\rm M}_{\sun})\geq9.2$.
It is interesting to compare the number density of RQEs to the simple
evolution model for massive ($>2.5\times10^{10}~h^{-2}~{\rm M}_{\sun}$) young
quiescent galaxies from \citet{whitaker12a}. 
These authors used the observed population
of old quiescent galaxies and modelled the build up of massive red-sequence
galaxies under the assumption that all such galaxies passed through
a brief young phase that lasted 0.5\,Gyr. 
Their prediction for $z=0.08$ is
$2.3\times 10^{-5}\,h^3\,{\rm Mpc}^{-3}$, compared to our lower (upper) limit of
$1.3(1.6)\times 10^{-5}\,h^3\,{\rm Mpc}^{-3}$
for RQEs using the same stellar mass cut. Therefore, we identify the bulk
(56--70\%) of the recently quenched galaxy population necessary to account
for the predicted red-sequence growth at late cosmic time.

In Table~\ref{tab:4},
we compare RQEs against two samples of non-star-forming ETGs:
(1) older blue, visually-selected Es and pEs from this work that
meet our modified $urz$ non-SF criteria and have
stellar ages $t_{\rm age}>3$\,Gyr; and
(2) red ETGs that meet our modified non-SF color criteria and have reliable
age estimates. Note, for both comparison samples we {\it do not} exclude
any emission types; i.e., these are simply color-selected and have good ages.
Using the half-light size and velocity dispersion, we estimate the
dynamical mass ${\rm M}_{\rm dyn}=5R_{50}\sigma^2/G$.
To explore the host environments of galaxies,
we use the SDSS DR4 galaxy group catalog. Details of the catalog's
construction, completeness and reliability are found in \citet{yang07}.
Briefly, the group catalog provides a number of physically motivated measures of
environment for all bright ($r\leq18$\,mag) galaxies in the NYU-VAGC
with redshifts $0.01\leq z\leq0.20$.
Here, we make use of (i)
a distinction between the highest stellar mass
galaxy (central$=$CEN) and less-massive group members (satellites$=$SATs);
and (ii) an estimate of the dark matter halo mass
${\rm M}_{\rm halo}$ of the host group, which is determined by ranking groups
in terms of characteristic stellar mass and applying halo occupation
statistics in the assumed $\Lambda$CDM cosmology. Using luminosity, rather
than stellar mass, to identify CENs and estimate halo masses does not affect
our results. Owing to the 
magnitude limit of the galaxies used to construct the catalog, groups
with $z\leq0.08$ are detected
with high completeness down to
$\log_{10}({\rm M}_{\rm halo}/h^{-1}{\rm M}_{\sun})=11.78$.

\begin{figure*}
  \center{
  \includegraphics[scale=0.65, angle=0]{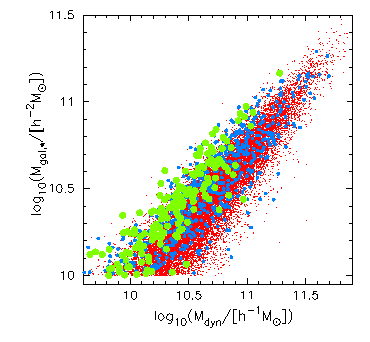}
  \hspace{0.3cm}
  \includegraphics[scale=0.65, angle=0]{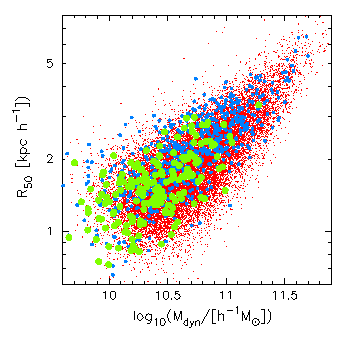}
  }
\caption[]{Stellar mass {\it left} and half-light size {\it right} 
of ETG samples from
Table~\ref{tab:4}) as a function of dynamical mass.
In each panel, we compare RQEs (green) to older ($t_{\rm age}>3$\,Gyr)
blue ellipticals (blue) and red ETGs (red) that both have $urz$ colors
within our modified non-SF regions (see text for detail).
The dynamical mass estimates are given by 
${\rm M}_{\rm dyn}=5R_{50}\sigma^2/G$.
The ${\rm M}_{\rm dyn}/{\rm M}_{{\rm gal,}{\star}}$ offset seen in the
RQE population may represent a systematic overestimate in 
the color-based stellar M/L ratios, which can be as much as 0.3--0.5\,dex
for a significant burst of recent SF \citep{bell01}. While the
dynamical-stellar mass mismatch can be unphysical for a few cases
(${\rm M}_{\rm dyn}<{\rm M}_{{\rm gal,}{\star}}$), the general trend is
consistent with the idea that RQEs are a post-starburst population that 
have unique SFHs compared to normal ETGs.
\label{fig:Mdyn}}
\end{figure*}

We find a number of interesting differences
between RQEs and the populations
of non-SF red ETGs and non-SF, older blue ellipticals.
While the three ETG populations in Table~\ref{tab:4} have similar stellar
mass and concentration distributions, a majority of RQEs have velocity 
dispersions and dynamical masses that are lower than
the median values for older ETGs. The difference in
${\rm M}_{\rm dyn}$ distributions is partially due to the fact that
RQEs are essentially a $<10^{11}~{\rm M}_{\sun}$ population while the
underlying ETG sample extends to higher masses. This highlights that RQEs
are the progenitors of a particular subset of ETGs.
But, we also see a clear ${\rm M}_{\rm dyn}/{\rm M}_{{\rm gal,}{\star}}$
offset when comparing RQEs and the bulk of ETGs in Figure~\ref{fig:Mdyn}.
At a fixed dynamical mass, RQEs typically have a 50\% larger stellar mass.
In some cases, the stellar masses are clearly unphysical with
${\rm M}_{{\rm gal,}{\star}}>{\rm M}_{\rm dyn}$.
This offset likely represents systematic errors in 
the color-based stellar M/L ratios, which can overestimate 
${\rm M}_{{\rm gal,}{\star}}$ by as much as 0.3--0.5\,dex
for a significant burst of recent SF \citep{bell01}. In this light, the 
majority of RQEs are consistent with a post-starburst population that is
unique compared to both red ETGs and older blue ellipticals in terms of
their SFHs. It is also possible that some of the dynamical-stellar mass 
differences reflect that the stars in different ETG populations
may probe different typical fractions of their dark halos.
Another possibility is that the RQEs could 
have a larger degree of rotation support
resulting in smaller measured velocity dispersions.
We note that one should take care in interpreting the sizes
of RQEs at fixed stellar mass compared to normal ETGs. 
In Figure~\ref{fig:Mdyn}, we show that RQEs have similar sizes as older
ETGs at a fixed dynamical mass.

Besides dynamical masses,
three quarters of RQEs reside in smaller groups than the medians of
both comparison samples. Nearly all RQEs are the dominant galaxy in
their host group, compared to 70\% for red ETGs. RQEs are
nearly four times more likely than red ETGs (and 50\% more likely than
older blue ellipticals) to house a Seyfert AGN and are much more
likely to have a post-starburst spectrum.
We discuss these properties in more detail with respect to quenching in
the next section.

Lastly, RQEs have a skewed
stellar metallicity distribution such that half of this sample
have values that are 50--100\% greater
than solar, in contrast to both non-SF comparison samples which have
median solar metallicities. 
One has to be careful in interpreting young galaxies with high metallicities 
as these follow the well-known age-metallicity relation \citep{worthey94b}.
Recently, several studies have argued 
that some anticorrelation between young ages 
and high metallicities may be real \citep{graves09b,trager09}, 
but the SDSS spectroscopic data are
too shallow to determine what fraction, if any, of the anticorrelation
is real. It will be worthwhile to pin down the metallicities of RQEs
since supersolar values may be an indication of metal enrichment from
a sustained period of enhanced SF prior to quenching. We note that
the lower metallicity nuclear gas found in tidally
interacting pairs locally \citep{kewley06a,ellison08c} is explained
by low-metallicity gas flowing from galaxy outskirts
during a major interaction \citep{rupke10a}, but quickly followed by chemical
enrichment from the subsequent SF \citep[e.g.,][]{torrey12}.
In this light, metal-rich RQEs would have distinct SFHs from the
rejuvenated, low-metallicity ETGs studied by \citet{thomas10}.

\begin{table*}
\caption{Properties of RQEs.}
\label{tab:4}
\begin{tabular}{l|c|cc}
\hline\hline
 &  & \multicolumn{2}{c}{Comparison Samples} \\
Property & Recently Quenched & non-SF, $>3$\,Gyr & non-SF\\
 & Ellipticals$^{\rm a}$ & Blue Ellipticals & Red ETGs\\
\hline
 $\sigma$ (km\,s$^{-1}$) & 99.4,\,122.6,\,149.2 & 115.6,\,143.4,\,179.3  & 128.1,\,156.4,\,191.5 \\
$\log_{10}({\rm M}_{\rm dyn}/h^{-1}{\rm M}_{\sun})$ & 10.20,\,10.41,\,10.64 & 10.40,\,10.67,\,10.94  & 10.45,\,10.68,\,10.93 \\
$\log_{10}({\rm M}_{{\rm gal,}{\star}}/h^{-2}{\rm M}_{\sun})$ & 10.23,\,10.39,\,10.58 & 10.29,\,10.50,\,10.72  & 10.24,\,10.43,\,10.64 \\
 $\log_{10}({\rm M}_{\rm halo}/h^{-1}{\rm M}_{\sun})$ & 11.86,\,12.06,\,12.36 & 11.95,\,12.30,\,12.77  & 11.95,\,12.38,\,13.24 \\
 $\log_{10}(Z/Z_{\sun})$  & 0.00,\, 0.23,\, 0.34 & $-$0.10,\, 0.01,\, 0.12  &  $-$0.04,\, 0.04,\, 0.13 \\
 $\log_{10}(t_{\rm age}/{\rm yr})$ & 9.20,\, 9.31,\, 9.39 & 9.64,\, 9.75,\, 9.85  & 9.81,\, 9.88,\, 9.93 \\
 $r$-band $R_{90}/R_{50}$ & 2.85,\, 3.02,\, 3.24 &  2.88,\, 3.06,\, 3.22  & 2.82,\, 3.00,\, 3.17 \\
\hline
 $\left \langle R_{50} \right \rangle$$^{\rm b}$ (kpc\,h$^{-1}$) & $1.61\pm0.34$ & $1.70\pm0.42$ & $1.50\pm0.38$ \\
\hline
 Number    & 172 & 651  & 23,094 \\
 central fraction    & 0.90  & 0.84   & 0.69  \\
 AGN fraction$^{\rm c}$ & 0.087(0.35)   & 0.055(0.29)  & 0.024(0.23) \\
 disturbed fraction  & 0.15  & 0.18   & na$^{\rm d}$ \\
 E$+$A fraction$^{\rm e}$ & 0.05 & 0.002 & 0 \\
\hline
\end{tabular}
\vskip 8pt
\begin{minipage}{\hdsize}
Properties of recently quenched ellipticals compared
against (1) older ($t_{\rm age}>3$\,Gyr), non-SF blue ellipticals, and
(2) non-SF red ETGs meeting the same 
${\rm M}_{{\rm gal,}{\star}}\geq 10^{10}\,h^{-2}{\rm M}_{\sun}$ 
selection. The comparison samples have $urz$ colors that fall
within our modified non-SF regions (see text for detail). 
For each galaxy sample, we tabulate
the 25\%, 50\% and 75\%-tiles of the distributions 
of velocity dispersion, dynamical mass,
stellar mass, halo mass, light-weighted stellar metallicity and age,
and $r$-band
central-light concentration. In addition, we list the mean half-light size
at the {\it same} fixed dynamical mass, plus we provide group center, 
AGN, disturbed morphology, and E$+$A fractions where applicable.\\
$^{\rm a}$ See \S~\ref{sec:QEidentif} for selection criteria.\\
$^{\rm b}$ Mean and standard deviation of the half-light size of galaxies
with dynamical masses within a narrow ($\pm20\%$) bin centered on
$2.5\times10^{10}\,h^{-1}{\rm M}_{\sun}$, the median ${\rm M}_{\rm dyn}$
for RQEs.\\
$^{\rm c}$ Seyfert (Seyfert $+$ BPT LINER $+$ \citet{yan06} weak-LINER) 
emission type fractions, see
\S~\ref{sec:EmTypes}.\\
$^{\rm d}$ Visual classifications were not done for red ETGs.\\
$^{\rm e}$ Based on the \citet{goto07b} catalog selected from SDSS DR5.
\end{minipage}
\end{table*}

\subsection{What Process Quenched Star Formation?}
\label{sec:QEhow}

As outlined in the Introduction, the growth of the red-sequence population
by the redward migration of blue-cloud galaxies requires both quenching of SF 
and the production of spheroid-dominated galaxies. By isolating RQEs
with the structure of red ellipticals, but the recent SFH of
blue galaxies, we can test for quenching processes that are both separate from
mechanisms that build spheroids such as environmental effects on gas,
and causally linked
to major structural changes such as major gas-rich merging. 
The recently quenched SF in these RQEs may have been triggered in some
part, or totally, by minor merging. Determining the amount of new
stars and the time since SF shutdown is beyond the scope of this study.
Instead, we focus on placing limits on the physics that
ultimately quenched the SF in RQEs.
In what follows,
we include energetic feedback processes which may or may not be directly
connected with merging, and
we exclude SF regulation processes that are predicted to occur only in disk
galaxies such as morphological quenching 
\citep[e.g.,][]{martig09} and bar-driven secular evolution 
\citep[e.g.,][]{masters11a}.

\subsubsection{Environmental Quenching}
The local environment of a galaxy can have a dramatic impact on its ability
to accrete cold gas and maintain the production of new stars.
Foremost, once the temperature of the gas in a galaxy's halo reaches the
virial temperature it will cool and accrete on to the galaxy with 
lower efficiency. Theoretical
calculations predict a relationship between the gas cooling time and the size
of a galaxy's dark matter halo \citep{rees77b,white78a}. Cosmological
and hydrodynamical simulations indicate that hot gas becomes the dominant medium
in halos with masses $\geq 10^{11}-10^{12}\,{\rm M}_{\sun}$ 
\citep{birnboim03,keres05}. Once hot gas
dominates, feedback mechanisms such as AGN heating may efficiently transfer
energy and momentum into the halo atmosphere and impede its subsequent accretion
on to galaxies that live in the halo
\citep{cattaneo06,croton06a,sijacki06}. 
Recently, \citet{dekel08} have calculated that deposition of gravitational
energy from cosmological accretion of dense cold gas clumps might 
heat halos without the need for AGN feedback.
Regardless of the heating source, hot halo quenching (i.e., impeding gas
cooling) is predicted to affect both
satellite (SAT) and central (CEN) galaxies \citep[e.g.,][]{gabor12}.
However, without additional feedback to quench the star-forming fuel supply
to the center of the halo, CENs in hot halos less massive than
${\rm M}_{\rm halo}\sim 10^{13}\,{\rm M}_{\sun}$ typically continue 
forming stars efficiently \citep{keres05,keres12,nelson13}.

The preferred environment of RQEs is intriguing in terms of the halo
quenching scenario. The majority of the RQEs we identify
are CENs in small dark matter halos (88\% have
$M_{\rm halo}<3\times 10^{12}\,h^{-1}{\rm M}_{\sun}$).
Therefore, if these galaxies are being quenched by a hot atmosphere in
their small host halo, we might
expect an increased incidence of AGN. We find that 7\% have optical
Seyfert emission and an additional 25\% have optical LINER emission.
If LINER emission and the typical AGN duty cycle are sufficient to
maintain halo heating,
then hot halo quenching may be the only mechanism required to explain
central RQEs in small groups. But, as mentioned earlier, 
much of LINER emission appears not to reflect AGN activity
and non-nuclear ionization sources are not reported to be energetic enough
to maintain halo quenching.
If, on the other hand, very efficient
AGN feedback from an active Seyfert is required to maintain quenching at the
centers of such small halos, then additional heating mechanisms would be
required if halo quenching is important for most RQEs. 
Even if quenching was initially triggered by an energetic outflow,
additional processes are still needed to prevent subsequent accretion and 
maintain long term quenching.
Besides AGN heating,
other options include gravitational heating \citep{dekel08} and SN feedback
\citep{springel03a}, but these are expected to work only in larger
($\geq 7\times 10^{12}\,{\rm M}_{\sun}$) and smaller
($<10^{11}\,{\rm M}_{\sun}$) halos, respectively, than the typical RQE hosts.
We note that we do not know for certain whether the host halos of RQEs are hot.
RQEs, especially those without optical
AGN emission, are good candidates for further testing of halo quenching
by looking for extended
hot halo gas using ROSAT stacking \citep[e.g.,][]{anderson13}.

The small SAT fraction suggests that satellite-specific processes are not
important for quenching most RQEs. It is possible that even fewer than
the small number (17) of these galaxies were not quenched as SATs.
First, some of the satellite RQEs may be interlopers.
Tests with mock redshift surveys show that the Y07 catalog has
an interloper fraction of $15\pm5\%$ \citep{yang05a}.
Such contaminants are typically CENs of low-mass halos along the line
of sight. We estimate the degree to which interlopers can enhance the RQE 
central fraction as follows. For the median stellar mass of RQEs,
\citet{vandenbosch08} found that the SAT fraction is roughly 25\%.
On the one hand, the interloper fraction implies that 3--5\% of galaxies
are actual additional CENs, which could raise the RQE central fraction
to as much as 95\%. On the other hand, applying the interloper fraction
to only the RQE satellites would add just 1--2\%.
Secondly, van den Bosch et al. calculated that the fraction of red-sequence
SATs that were {\it quenched as satellites} drops from 30\% at
${\rm M}_{{\rm gal,}{\star}}=10^{10}~h^{-2}~{\rm M}_{\sun}$ to a few
percent at the maximum RQE mass. Assuming that these results apply to
the unique subset of RQEs, it is possible that any legitimate SATs
might have quenched prior to falling into their host halo; i.e., it is
possible that {\it all} RQEs were quenched as CENs.

Assuming that some of the RQEs are legitimate SATs, the stripping
of the hot gas atmosphere of the galaxy (strangulation) is argued to be both
the main quenching mechanism for SATs and efficient at all halo masses
\citep{vandenbosch08}. With regard to ram-pressure stripping, only
one-quarter of the satellite 
RQEs reside in cluster-sized halos (masses larger than 
$\sim 10^{14}\,h^{-1}{\rm M}_{\sun}$) in which it is expected that
rapid motion of the SAT through the ICM can remove gas.
Moreover, these SATs all lie at large projected
cluster-centric distances ($>300$\,kpc) where ram-pressure stripping would be
least effective.
We note that more than half of the SATs are found in 
$>10^{13}\,h^{-1}{\rm M}_{\sun}$ halos, which is consistent with the
masses needed for
hot halo quenching (impeding gas cooling). 
If this is the responsible quenching mechanism, an
open question is when were these previously star-forming 
ellipticals added to their halo -- before or after quenching? 
If the quenching event marked the halo entry, this would support the idea
that the gas accretion rate of a SAT declines rapidly (0.5--1\,Gyr) after
it enters a large halo \citep{simha09}. Conversely, \citet{wetzel13} have
argued for a `delayed-then-rapid' quenching scenario in which
SATs spend 2--4\,Gyr forming stars at a normal rate after initial infall,
before finally rapidly quenching.

We note that the satellite RQEs have a much higher 
Seyfert fraction (albeit small numbers: five out of 17 $\approx 30\%$) 
and a similar LINER fraction compared to their
CEN counterparts. In both cases, the incidence of such activity in SATs 
does not depend
on environmental measures such as halo mass or distance from halo center
in agreement with the behavior of the general SAT population \citep{pasquali09}.
It is interesting to speculate whether some of these SATs,
especially the Seyferts,
are recent merger remnants that actually mark the centers of substructure in 
their host group; e.g., \citet{mcintosh08} argued that SAT-SAT mergers may
represent the dynamical centers
of subhalos that have recently accreted on to their host.
In fact, \citet{simha09} showed that mergers between satellite galaxies 
within infalling substructures are common in cosmological simulations.
 
\subsubsection{Quenching by Energetic Feedback}
\label{sec:Qagn}
The addition of energy released from supermassive black hole accretion
(AGN feedback) or from super novae and massive-star winds during 
a burst or strong
enhancement of SF (SN or stellar feedback) are argued to be important
quenching mechanisms. For example, \citet{kaviraj11b} make the case
for AGN feedback arguing that
gas depletion by SF alone is too slow to explain the blue-to-red
migration rate observed in local ETGs by \citet{schawinski07b}.
Compared to `radio mode' AGN feedback,
which simply keeps the atmospheres of groups and clusters
hot as discussed in the previous section, simulations show that
energetic or mechanical AGN feedback can remove and/or
dynamically disrupt the gas resevoirs of isolated elliptical
galaxies \citep[e.g.,][]{gaspari12}.
Calculations show that nuclear inflows
of gas are caused by non-axisymmetric pertubations from both galaxy
interactions \citep{barnes96,mihos96,springel05a}
and from large-scale stellar bars \citep{shlosman89}, which provides
theoretical grounds for the debate in the literature about whether a 
major merger-AGN connection exists or not
\citep[e.g.,][]{cisternas11}. Therefore,
here we first discuss the evidence for feedback in our sample of RQEs
regardless of the triggering mechanism, and in the next section we investigate
whether any connections can be made with recent major merging.

We noted the incidence of optical Seyfert and LINER emission in the previous 
section.
As Table~\ref{tab:4} shows, Seyfert activity is more than three times higher
in RQEs compared to red and dead ETGs. Conversely, there is no difference
in the LINER fraction. Including LINERs, the optical AGN fraction (35\%)
for RQEs is much larger than the 8\% quoted by \citet{mendel13} using the
{\it same} BPT criteria for a larger sample of recently
quenched SDSS galaxies. We conclude that this difference comes from their
sample selection which includes galaxies that are up to $\sim6$ times
less massive than our low-mass cutoff. An increased AGN fraction and, hence,
importance with
higher stellar mass may seem natural at first given the fundamental
relationship between black hole and galaxy spheroid masses
\citep{ferrarese00,gebhardt00}. The stellar mass dependence of AGN incidence
has been linked to SF quenching efficiency to argue
that AGN feedback dominates at
${\rm M}_{{\rm gal,}{\star}}>10^{10}~{\rm M}_{\sun}$ \citep{kaviraj07d}.
But, \citet{aird12} argue that the distribution of AGN accretion rates
are independent of the host galaxy mass and find no direct relation between
the presence of an AGN and quenching. 

Of all the RQEs, the active Seyferts provide an association
between nuclear activity and recent quenching.
Our finding is in qualitative agreement with the maximum Seyfert fraction
found by S07 and \citet{schawinski10a} 
among galaxies in the green valley that 
are presumably migrating redwards.
We note that the Seyfert incidence in RQEs is more than 60\% higher
than in similarly non-star-forming, but older, blue ellipticals.
Yet, Seyfert lifetimes are very brief 
\citep[$\leq 100$\,Myr,][]{martini04,hopkins06f} compared to the
RQE phase (presumably $\sim$ A-star lifetime), thus, the observed
activity in RQEs is likely unrelated to the actual quenching.
It is possible that these Seyferts are the later stage of much more powerful
obscured QSO phases \citep{hopkins05d} that quenched SF, or they are
merely the result of some gas remaining in the nuclei of some RQEs.
Likewise, some LINERs might be the very tail end of the quenching
AGN as schematically illustrated in Fig.\,1 of \citet{hopkins08a},
or they may simply reflect the underlying
fraction of ETGs with LINER emission
\citep[many of which are not associated with nuclear ionization; e.g.,][]{annibali10}.
The simple fiber emission-line diagnostics we use do not allow us to
explore these possibilities. Potentially, 
a direct AGN feedback link can be demonstrated with observations of 
high-velocity
outflows tied to an AGN in systems that have recently quenched.
So far, observations of Seyferts with outflows are quite limited at low redshift
\citep{riffel09,mullersanchez11,crenshaw12},
and totally lacking for recently quenched galaxies.
Assuming that the RQE-Seyfert association reflects the frequency of actual
AGN feedback, then the low Seyfert fraction would imply that this
mechanism is not important for the majority of high-mass RQEs.

With regard to stellar feedback, we find a remarkably significant, yet
low, percentage (5\%) of RQEs that are identified as E$+$A by
\citet{goto07b}. Our fraction is in good agreement with that found by
\citet{mendel13} in recently quenched galaxies extending to lower stellar mass.
This percentage is significant compared to older non-SF blue ellipticals (0.2\%),
red non-SF ETGs (0\%), and the total SDSS galaxy sample
\citep[0.1\%,][]{goto07b,wong12}. The interpretation of the
E$+$A signature has long been a large starburst that was rapidly and
recently truncated \citep{dressler83,chilingarian12}.
But, it remains unclear if the burst was responsible for quenching the RQE
in the E$+$A subset. Indeed, all of the RQEs are post `starburst' at some
level given their unusually blue colors but lack of SF. Our results suggest
that most quenching does not involve a starburst big enough to make a 
long-lived E$+$A signature. Under the assumption that stellar feedback is
important only in strong post-starburst systems, then the weak E$+$A association
would imply that stellar feedback is not important for the majority of
high-mass RQEs. Moreover, if all RQEs (or a large fraction) are examples
of first-generation ellipticals, then E$+$A studies are exploring just
a small subset of the newly-formed elliptical population.

\subsubsection{Quenching by Major Merging}

The work of Hopkins, in particular, has given a lot attention to the
gas-rich major-merger production of red ellipticals as the natural link between
hierarchical evolution and the migration of galaxies necessary to explain
the buildup of the red sequence. The merger-fueled formation of
new stars followed by efficient quenching of SF is a compelling
model that naturally explains blue-to-red color evolution, the destruction of
disks, and the formation of elliptical galaxies.
Indeed, with sufficiently high gas fractions, this model nicely ties together
the evolution of starburst and quasar activity with the production of
galaxy spheroids over cosmic time
\citep{hopkins06b,hopkins08b}.
While the cold gas fractions in present-day disk galaxies are much lower
than at $z\sim1$, major
merging remains the best way to form new elliptical galaxies
and a viable way to quench recent SF. In this context, the RQEs of this study
provide an important elliptical galaxy subpopulation with which to
test the modern merger hypothesis since it predicts a RQE phase in which the
formation of an elliptical and the quenching of SF are clearly causally linked.
Here, we discuss to what degree RQEs agree with the major merger scenario
and test the plausibility of a major merger-quenching connection by
exploring their disturbed fraction, AGN fraction, and preferred environment.
We then compare the frequency of RQEs to that of local gas-rich mergers.

{\bf Test 1: Morphological Disturbance.}
An obvious test for a merger-RQE connection is signs of recent tidal activity.
Not withstanding our uncertainty in recognizing morphologically peculiar
ellipticals, the low disturbed fraction (15\%) among RQEs,
naively, may conflict with idea that all RQEs are made in major mergers. 
There are two possible interpretations (where both may be at play): 
not all RQEs are made in major mergers, or the time-scales for 
identifying post-merger 
disturbed morphology are substantially shorter than the time-scale 
over which the stellar population signatures of RQEs are seen.

It is noteworthy that RQEs do not show an enhanced disturbed fraction
over older, non-SF blue ellipticals (see Table~\ref{tab:4}).
As \citet{feldmann08} showed, ellipticals with clear morphological
disturbances (pE types) could be the result of recent accretion of a minor
satellite on to a pre-existing elliptical. 
Distinguishing new end products of minor and major merging is beyond
the scope of this study. If any disturbed RQEs are the result of minor
accretion, these galaxies would be consistent with frosting of an existing
elliptical galaxy and inconsistent with major merger quenching.
Unfortunately, morphology alone does not provide conclusive tests
of the merger hypothesis. Multiple studies find a broad association 
between peculiar morphologies and younger populations 
\citep{schweizer92,tal09,gyory10}, which lends support to the notion that 
recent SF is often associated with some kind of interaction (either
minor or major), but the scatter is large in the observed correlations.
In this light, the level of morphological disturbances among RQEs is consistent 
with the scatter in the age-asymmetry correlations.

If we assume that every RQE was formed by a recent
major merger of spiral galaxies, the
low fraction of obvious (i.e., strong) tidal features
may simply provide information about the dynamical age of these galaxies (i.e.,
time since coalescence). Merger simulations predict that major gaseous
disk-disk mergers produce spheroidal remnants 
that start highly disturbed (qualitatively like our SPM classification),
and then dynamically relax and settle into a normal elliptical
morphology over time.
As with efficiency of merger-induced SF \citep{cox08a},
the post-merger morphology and its time-scale to relaxation
depends on the suite of merger parameters, but typically the strong
morphological disturbances are short-lived and insensitive to viewing angle
and most merger orbits. For example, strong asymmetries
fade in $<0.5$\,Gyr in realistic images generated from a series of simulated
disk-disk mergers spanning {\it major} mass ratios and present-day massive
spiral gas fractions
\citep{lotz08b,lotz10a,lotz10b}. 
Using deeper SDSS Stripe 82 data, \citet{schawinski10b} argued that
major morphological disturbances in green-valley ETGs decline rapidly
(in several hundred Myr) after a merger-driven starburst to the levels
of quiescent ETGs. 
It would be interesting to observe RQEs with
deeper imaging to look for evidence of fairly recent tidal activity.

{\bf Test 2: AGN Fraction.}
Under the right conditions (e.g., orbital configuration, mass ratio, gas
fraction, progenitor morphology), 
hydrodynamical simulations demonstrate that disk-disk merging 
produces radial inward gas flows that fuel central SF and
supermassive black hole growth 
\citep{barnes91,barnes96,mihos94b,mihos96,cox08a,johansson09a}.
As such, energetic AGN and SN feedback are the favored quenching mechanism
in new major merger remnants \citep{springel05a,cox06b,khalatyan08}. In this
light, the fact that
RQEs have the highest incidences of Seyfert emission (9\%) and
post-starburst signatures (5\%) among local ETGs supports the idea that
some fraction of this population formed in a gas-rich merger that
triggered their quenching. If we assume that the
RQEs with LINER emission were formed by a recent merger and we are 
witnessing the low-power tail end of merger fueled AGN activity,
then 35\% of RQEs have spectroscopic signatures that are
consistent with energetic feedback quenching triggered by a merger.
Yet, this merger-AGN connection is tenuous given that a fifth of non-SF
red ETGs are LINERs (see Table~\ref{tab:4}), and given the fact that
many LINERs may not be AGNs. On the other hand,
\citet{hopkins08b} argued that major merging could provide at least
temporary quenching purely by shock heating the gas, but it is not clear if
this would work over very long time-scales. Under the assumption that RQEs
are the result of a recent merger, it is possible that this temporary
quenching could account for the majority of the RQEs with no AGN
\citep[but others argue AGN feedback is mandatory; e.g.,][]{springel05a}.
In this light, optical AGN fractions do not provide conclusive interpretations
with respect to major merging quenching \citep[see also,][]{schawinski10b}.

{\bf Test 3: Environment.}
We find that the preferred environment of RQEs agrees well with 
hierarchical merging predictions. As discussed in the previous sections,
90\% of RQEs reside at the centers of their host halo and these host halos
are typically small with a median mass of
$\log_{10}({\rm M}_{\rm halo}/h^{-1}{\rm M}_{\sun})=12.03$.
\citet{hopkins08a} defined the `small group scale' to describe the lowest-mass
halos with the highest probability to host a pair of similar mass galaxies,
and showed that major mergers are most likely to occur at the centers of
such halos. Further, they calculated the halo mass in which major 
merging on to the central galaxy is most efficient as a function of
galaxy mass (see their Fig.\,4).
We compare these calculations to the typical halo masses of central RQEs,
under the assumption that these galaxies formed in a recent spiral-spiral
merger at the center of their host halos.
For example, the median stellar mass central RQE has
a mass of $2.6\times 10^{10}~h^{-2}~{\rm M}_{\sun}$ and lives in a
$\log_{10}({\rm M}_{\rm halo}/h^{-1}{\rm M}_{\sun})=12.02$ halo. 
If this galaxy formed
by a major 1:3 (or 1:1) merger of a 
$6.5\,(13) \times 10^{9}~h^{-2}~{\rm M}_{\sun}$ galaxy on to the central
object, then Hopkins et al. predict a 
$\log_{10}({\rm M}_{\rm halo}/h^{-1}{\rm M}_{\sun})=12.0$ (12.2)
halo to have the maximum
efficiency for such a merger when we extrapolate using a characteristic mass
for LTG's (spirals) of $\log_{10}({\rm M}^*/h^{-2}{\rm M}_{\sun})=10.5$
from the local Schechter mass function of \citet{bell03b}.
Likewise, for central RQEs in halos spanning the upper and lower quartiles
of the mass distribution,
the peak merger efficiency occurs in halos with 
small group scale masses between
$11.9<\log_{10}({\rm M}_{\rm halo}/h^{-1}{\rm M}_{\sun})<12.4$,
which is in very good agreement with the RQE percentiles
$11.8<\log_{10}({\rm M}_{\rm halo}/h^{-1}{\rm M}_{\sun})<12.3$.

{\bf Test 4: Comparison to Merger Demographics.}
As a final test, we compare the incidence of RQEs to
the demographics of
major spiral-spiral interactions at low redshift.
This comparison provides some interesting insights
into the likelihood of mergers that produce starbursts, AGNs and 
quenching in accord with simulations. Using results
from a forthcoming analysis of the same complete selection of
high-mass SDSS galaxies, we find that the number density of
major interactions between two star-forming galaxies based on the H12 $urz$
criteria is 65\% to twice that of RQEs (the range represents the strength
of visible disturbances in the galaxy pairs, McIntosh et al., in prep.). 
If we select the lower limit of spiral pairs with the strongest interaction
signatures, and assume that these disturbances are visible for half the
typical RQE phase time-scale,
then this result suggests that all RQEs could have been formed in such
a merger, recently.

\subsection{Plausible Descendants of RQEs}
\label{sec:descendants}

As described in \S~\ref{sec:QEnature},
the nature of RQEs is unique compared to typical {\it red-and-dead} ETGs.
In the context of the expected redward evolution of RQEs, it is worthwhile
to ask if any red galaxies with similar properties but older stellar ages
(i.e., plausible descendants) exist at late cosmic times.
To answer this question, we search the red-sequence population of ETGs
for older non-star-forming galaxies
of similar small velocity dispersion and high stellar 
metallicities as the RQEs. As shown in Table~\ref{tab:4}, the velocity
dispersion (and related dynamical mass) and
metal richness of RQEs are the most striking differences in comparison
to typical non-SF red ETGs, aside from AGN and E$+$A fraction differences.
Under the simple assumption that the RQEs will experience no new SF,
the young stars will evolve and we expect
that their descendants should maintain their mass and size as they
evolve redwards as long as they do not have many minor mergers in
the interim which are predicted to puff up ETG sizes \citep{bezanson09}. 
Clearly, stellar aging will reduce the frequency of systems
with post-starburst signatures to zero, and we expect a likewise drop
in Seyfert systems since this activity is predicted to be a brief phase,
especially if it is tied to merger-driven AGN feedback
\citep[e.g.,][]{hopkins08a}.

With the above constraints in mind,
we focus our search on red ETGs with the highest
$\log_{10}(Z/Z_{\odot})$ values
(we use the upper metallicity quartile for non-SF red ETGs,
see Table~\ref{tab:4}),
and velocity dispersions
that fall within the upper and lower quartiles of the RQE population.
Additionally, we require plausible RQE descendants to have light-weighted
ages $>3$\,Gyrs, meet the H12 non-SF criteria, be similarly round
($b/a>0.6$) as typical RQEs, and have AGN (LINER or Seyfert) emission
or be spectroscopically quiescent; i.e., no SF or Comp emission types.
These criteria result in 475 plausible descendants with
$99.4\leq \sigma \leq 149.2$\,km\,s$^{-1}$ and
$\log_{10}(Z/Z_{\odot})>0.13$. Thus, we find that
low-$\sigma$, round, metal-rich ETGs
with older stellar ages do exist at low redshift. 

The majority of plausible RQE descendants and RQEs with matched velocity
dispersions have
comparably small sizes ($R_{50}<2\,h^{-1}$\,kpc). 
With regard to light-weighted stellar metallicities, we find that
the most metal-rich descendant candidates have similar (somewhat lower but
within the quoted $\pm 0.15$\,dex uncertainties) 
metallicities as RQEs. The metal-rich
descendants make up the upper 10\% of similar low-$\sigma$ red ETGs, which
tend to have closer to solar metallicities; i.e., such low-mass, metal-rich
ETGs exist, but they are fairly unusual compared to the fraction of
metal-rich systems among higher-mass ETGs.
In \S~\ref{sec:QEnature}, we discussed the issues with 
interpreting high metallicities in light of the
age-metallicity degeneracy \citep{worthey94b}. Nevertheless, if we take
the RQE and descendant candidate metallicities at face value,
we speculate that small differences in median metallicities between these two
populations can arise simply from aging of recently formed metal-rich stars
and the subsequent decrease of their contribution to the overall luminosity.
In terms of basic structural properties, this population of plausible
RQE descendants is consistent with
evolved RQEs and has a number density of
$7.4\times 10^{-5}\,h^3\,{\rm Mpc}^{-3}$ that is 5.5 times greater
(at matched $\sigma$); i.e., there are enough of these galaxies to
account for the expected color evolution of RQEs.

We investigate the environments of the plausible descendants 
and find that the fraction
that reside at the center of their host group
is only 64\%. This value is somewhat lower than for all non-SF 
red ETGs, and much lower than the 93\% central fraction for the RQEs
with matched velocity dispersions. The central RQE descendants 
reside in small groups with an average 
halo mass of $1.2\times 10^{12} h^{-1}{\rm M}_{\sun}$, which is comparable
to the size of halos with a central RQE. Therefore, given the larger
number density of plausible descendants, every central RQE
could evolve on to the red sequence at the center of its host halo 
and be consistent with the 64\% CEN fraction. 
With regard to quenching, we find that the CEN descendant candidates
have high AGN and Seyfert fractions of 51\% and 5\%, respectively.
Given their median age ($t_{\rm age}=6$\,Gyrs), these fractions are
noteworthy for two reasons. First, it appears that Seyferts can occur
in small red ellipticals presumably long after SF was quenched, which
calls into question connections between short duty cycle AGNs
and blue-to-red migration (e.g. S07).
Secondly, the high fraction (46\%) of LINER activity in these CENs
is consistent with the idea that heating from a low-power AGN
\citep{cattaneo06,croton06a,sijacki06}
could be keeping SF shut off at the centers of these small halos,
which would otherwise accrete enough gas to remain star-forming 
\citep{keres05,keres12,nelson13}. Whether the typical
ionization source of LINERs is post-AGB stars or 
a low-accretion black hole, it would be interesting to test if these
objects can provide sufficient heating in small dark halos.

\section{Summary}
\label{sec:Summary}

We analyze a large sample of high-mass, ETGs from
the SDSS DR4 that
are unusually blue; i.e., bluer than the empirical red sequence with
typically green-valley colors. This sample is volume-limited
($0.01<z\leq 0.08$) and stellar mass-limited
(${\rm M}_{{\rm gal,}{\star}}\geq10^{10}~h^{-2}~{\rm M}_{\sun}$).
Through careful morphological inspection we identify an important subsample of
1500 blue elliptical (pure spheroid) galaxies; nearly 10\% have 
morphological peculiarities that are
suggestive of recent tidal activity (either major or minor merging).
These bluest ellipticals comprise 3.7\% of high-central concentration
selected ($c_r\geq 2.6$) ETGs with stellar masses between
$10^{10}$ and $10^{11}~h^{-2}~{\rm M}_{\sun}$.
The vast majority (95\%) of
visually-selected blue ellipticals have axial ratios of $b/a>0.6$, which
is a cut that \citet{zhu10} employed to remove S0 galaxies. Moreover, these
blue galaxies follow a stellar mass-size relation that
is consistent with red-sequence ETGs and {\it distinct} from star-forming 
disk galaxies. 
Very few blue ellipticals have stellar masses in excess of
${\rm M}_{{\rm gal,}{\star}}=10^{11}~h^{-2}~{\rm M}_{\sun}$, which is
consistent with the idea that dry merging dominates the mass assembly
of the most massive
ellipticals \citep[e.g.,][]{mcintosh08,vanderwel09b}.
In terms of nebular emission, we find
that unusually blue ellipticals are quite active with
distinct emission fractions compared to three control samples:
red ETGs, blue disk-dominated galaxies, 
and blue ETGs with visual disk features
(i.e., bulge-dominated spirals and S0 galaxies).
The incidence of blue ellipticals with optical SF emission is much greater
than that of red ETGs, but lower than for both blue disk galaxy populations.
Blue ellipticals are 50\% more likely to have strong emission
(S/N$\geq3$ in all four BPT lines) than red ETGs, yet 40\% are
spectroscopically quiescent. We divide
blue elliptical galaxies by optical emission into several subpopulations
in Figure~\ref{fig:colorSigma} and find that they
follow a similar sequence of emission activity in the color-sigma plane as S07
found for a broad selection of ETGs not limited by visual classification,
color or stellar mass. We find that the subset of disturbed blue ellipticals
prefer to populate the bluer, high-velocity-dispersion edge of this 
distribution for all emission types except Seyferts.
This result is tentative given possible identification biases at the lowest
masses, but it is also intriguing in light of scenarios that involve
merging and quenching to explain the mass limit of the blue cloud
\citep[e.g.,][]{faber07}.

While high-mass, blue elliptical galaxies with specific emission properties
(e.g., star-forming ellipticals) are interesting in their own right,
here we focus on one subset,
a unique population of RQEs.
We find that many quiescent blue ellipticals have distinct 
$(u-r)$--$(r-z)$ colors 
(Fig.~\ref{fig:urz_RQE_seln}) that extend bluewards of the robust
selection of non-star-forming galaxies by H12, and are very
young based on \citet{gallazzi05} light-weighted stellar age estimates.
As such, we confirm a similar color-based selection of young quiescent
ETGs at $z>1$ by \citet{whitaker12a}.
We use modified non-star-forming $urz$ colors, 
$t_{\rm age}<3\,$Gyr, and a lack of detectable emission from SF to identify
172 RQEs. These are a unique subpopulation of ETGs; i.e., virtually zero
red-sequence ETGs have the ages {\it and} colors of RQEs.
These galaxies clearly experienced a recent quenching of SF
and are now transitioning to the red sequence. 
We argue that RQEs have recent SFHs that are distinct from similarly young and
blue ETGs with ongoing SF \citep[i.e., rejuvenated ETGs,][]{thomas10}.
Compared to older blue and red non-SF ETGs, 
RQEs are offset to larger stellar masses which likely reflects 
stellar M/L ratio overestimates from a recent burst of SF \citep{bell01}.
Given their unusually blue colors, all RQEs are post `starburst' at some
level. While we find a
significant incidence (5\%) of RQEs identified as post-starburst by
\citet{goto07b}, we conclude that
most quenching does not involve a starburst big enough to produce 
an E$+$A signature.
Keeping the caveat in mind that one must take care in interpreting 
young galaxies with high metallicities
in light of the age-metallicity degeneracy, we note that
the majority of RQEs have Gallazzi et al. stellar metallicity estimates
above solar, unlike similar mass rejuvenated ETGs. Taken at face value,
metal-rich RQEs are consistent with chemical enrichment from a significant
merger-triggered SF event prior to quenching \citep{torrey12}.
Given the expected redward evolution of RQEs, we identify
similarly low-dynamical mass, metal-rich, and non-star-forming ETGs
with older stellar ages (plausible RQE descendants) at low redshift 
with sufficient numbers to account for these RQEs.

RQEs are strong candidates for `first-generation' ellipticals formed in a
relatively recent major spiral-spiral merger, thus, RQEs provide a useful
test of the modern merger hypothesis for the origin of new ellipticals
\citep[e.g.,][]{hopkins08b}. 
A number of RQE properties are consistent with a recent gas-rich merger
origin. Comparing the incidence of RQEs to spiral-spiral major-merging
demographics from the SDSS (McIntosh et al., in prep.) suggests
that all such mergers can produce a RQE under the simple assumption
that the RQE time-scale is twice the tidal interaction visibility time-scale.
In terms of environmental metrics from \citet{yang07}, 
RQEs reside in smaller groups 
(90\% in $M_{\rm halo}\leq 3\times 10^{12}\,h^{-1}{\rm M}_{\sun}$)
than red ETGs, and are the most massive (`central')
galaxy in their host halo more often (90\% vs. 70\%).
The preferred environment of RQEs agrees well with
hierarchical merging predictions of \citet{hopkins08a}, such
that most reside at the centers
of small dark matter halos with masses that match the
small group scale in which spiral merging on to the halo center is 
maximally efficient. We argue that
the low (15\%) disturbed fraction is consistent with the large scatter in
the observed age-asymmetry correlation \citep{schweizer92,tal09,gyory10},
and likely reflects that the time-scale for
identifying post-merger tidal signatures is substantially shorter than
the RQE phase lifetime.

The number density of RQEs is $2.7(4.7)\times 10^{-5}\,h^3\,{\rm Mpc}^{-3}$;
the upper limit includes blue ETGs with uncertain
classifications that may be elliptical in morphology.
For masses above $2.5\times10^{10}~h^{-2}~{\rm M}_{\sun}$, 
the number density is comparable to the simple evolution model from
\citet{whitaker12a}, such that massive RQEs can account for 
56--70\% of the young quiescent galaxies necessary to match the quiescent
red-sequence growth at late cosmic time assuming a 0.5\,Gyr time-scale
for the RQE phase. As such, RQEs provide a testbed for
constraining different quenching processes related to {\it one}
(possibly the most important) {\it channel}
of blue-to-red migration -- the addition of new ellipticals
to the red sequence.
If we assume that RQEs were formed by gas-rich mergers,
then the higher incidences of Seyfert emission (9\%) and E$+$A signatures
compared to the bulk of local ETGs are consistent with the idea that
some fraction of this population was quenched relatively recently
by merger-triggered energetic feedback. But this fraction is small, even
if one assumes that Seyfert activity is the tail end
of a more powerful QSO and stellar feedback dominates only in
strong post-starburst systems.
These associations are mainly circumstantial and the emission-line
diagnostics we employ do not allow us 
to make a direct energetic feedback-quenching
(either AGN or SN) nor merger-quenching connection for RQEs. 

Instead, our analysis allows us to rule out satellite-specific quenching
processes for 90\% of RQEs
and provides a number of constraints on likely quenching mechanisms for
most RQEs.
Foremost, the high fraction of central RQEs in small
($<3\times 10^{12}\,{\rm M}_{\sun}$) halos indicates
that {\it if} these galaxies were quenched by the inability of the hot halo 
atmosphere to cool, then
additional feedback mechanisms are needed to maintain the heating and
halo quenching \citep[e.g.,][]{keres05}.
Even if quenching was initially triggered by a very energetic process 
(e.g., strong AGN feedback), additional processes are still needed to 
maintain long term quenching.
Depending on the assumed duty cycle, the incidence of Seyferts and LINERs
may be sufficient to maintain halo heating. While the LINER fraction of
RQEs is no different from that of old red ETGs, 
we find a significant excess ($\sim 50\%$) in the LINER incidence among
the plausible {\it central} RQE descendants residing in similar mass halos,
which supports the low-power AGN heating model for keeping SF shut off
long after it was initially quenched.
But, if LINERs (regardless of the
ionizing source) are not energetic enough to effectively heat these halos,
then the low frequency of Seyfert activity would imply that 
additional heating mechanisms must be at play if halo quenching
is important for the majority of RQEs. This opens the possibility for
unexplored heating processes or for known sources
(e.g., SN feedback or gravitational heating) that are
effective at halo masses different from what is predicted.
Further analysis of RQEs and their likely descendants will provide
valuable additions to our understanding of the complex processes that
govern galaxy evolution.

\section*{Acknowledgements}
We thank the anonymous referee for useful 
comments that strengthened this paper.
We are grateful for useful discussions with Alison Coil, St\'{e}phane Courteau,
T.\,J. Cox, Lars Hernquist, Phil Hopkins, Neal Katz, Hans-Walter Rix, Greg Rudnick, Arjen van der Wel
and Martin Weinberg during different manifestations of this work.
D.H.M., C.W., A.C., and J.M. acknowledge support from the Research 
Corporation for Science Advancement under the Cottrell College Science Award
grant No. 10777. C.W., A.C., and J.M. acknowledge support from the Missouri
Consortium of NASA's National Space Grant College and Fellowship Program.
A.G. acknowledges support from the EU FP7/2007-2013 under grant 
agreement No. 267251 AstroFIt.
This publication makes use of the Sloan Digital Sky Survey (SDSS).
Funding for the creation and distribution of the SDSS Archive has been 
provided by the Alfred P. Sloan Foundation, the Participating Institutions, 
the National Science Foundation, the U.S. Department of Energy, 
the National Aeronautics and Space Administration, the Japanese Monbukagakusho, 
the Max Planck Society, and the Higher Education Funding Council for England. 
The SDSS Web Site is http://www.sdss.org/.
The SDSS is managed by the Astrophysical Research Consortium for the 
Participating Institutions. The Participating Institutions are the
American Museum of Natural History, Astrophysical Institute Potsdam, University of Basel, University of Cambridge, Case Western Reserve University, University of Chicago, Drexel University, Fermilab, the Institute for Advanced Study, the Japan Participation Group, Johns Hopkins University, the Joint Institute for Nuclear Astrophysics, the Kavli Institute for Particle Astrophysics and Cosmology, the Korean Scientist Group, the Chinese Academy of Sciences (LAMOST), Los Alamos National Laboratory, the Max-Planck-Institute for Astronomy (MPIA), the Max-Planck-Institute for Astrophysics (MPA), New Mexico State University, Ohio State University, University of Pittsburgh, University of Portsmouth, Princeton University, the United States Naval Observatory, and the University of Washington.
This publication also made use of NASA's Astrophysics Data System
Bibliographic Services and TOPCAT \citep[Tools for OPerations on Catalogues And 
Tables,][]{taylor05}.

\footnotesize{
  \bibliographystyle{mn2e}
  \bibliography{/Users/danmcintosh/Papers/refs4mn2e.052214.bib}
}

\bsp
\label{lastpage}

\end{document}

%% file: macros.tex
\newdimen\hssize 
\newdimen\hdsize 